\documentclass[12pt]{article}
\usepackage{amsmath,amssymb,graphicx,bbm}
\usepackage[utf8]{inputenc}
\numberwithin{equation}{section}
\usepackage{color}
\usepackage[dvipsnames]{xcolor}

\usepackage[linktocpage=true]{hyperref}
\hypersetup{
  colorlinks=true,
  linkcolor=blue,
  citecolor=blue,
  pdfpagemode=FullScreen,
  }

\edef\restoreparindent{\parindent=\the\parindent\relax}
\usepackage{parskip}
\restoreparindent

\usepackage{tikz}
\usetikzlibrary{arrows,arrows.meta,intersections, calc,positioning,decorations.pathreplacing,decorations.pathmorphing,shapes}
\usetikzlibrary{patterns}
\usetikzlibrary{decorations.markings}
\usetikzlibrary{knots}

\usepackage{here}
\usepackage{caption}
\begin{document}
\begin{titlepage}

\renewcommand{\thefootnote}{\fnsymbol{footnote}}
\begin{flushright}
\begin{tabular}{l}
YITP-25-13\\
\end{tabular}
\end{flushright}

\vfill
\begin{center}

\noindent{\Large \textbf{AdS gravastar and bulk-cone singularities}}



\vspace{1.5cm}

\noindent{Heng-Yu Chen,$^{a,c}$\footnote{heng.yu.chen@phys.ntu.edu.tw} Yasuaki Hikida$^{b,c}$\footnote{yhikida@yukawa.kyoto-u.ac.jp} and Yasutaka Koga$^{b,c}$\footnote{yasutaka.koga@yukawa.kyoto-u.ac.jp}}

\bigskip

\vskip .6 truecm

\centerline{\it $^a$Department of Physics, National Taiwan University, Taipei 10617, Taiwan }

\medskip

\centerline{\it $^b$Faculty of Information Science and Technology,}
\centerline{\it  Osaka Institute of Technology,}
\centerline{\it Kitayama, Hirakata, Osaka 573-0196, Japan}

\medskip

\centerline{\it $^c$Center for Gravitational Physics and Quantum Information,}
\centerline{\it  Yukawa Institute for Theoretical Physics, Kyoto University,}
\centerline{\it Kitashirakawa Oiwakecho, Sakyo-ku, Kyoto 606-8502, Japan}

\end{center}

\vfill
\vskip 0.5 truecm
\begin{abstract}

The horizon of black hole is surrounded by the photon sphere and an outside observer cannot easily examine the geometry inside the photon sphere. In this note, we propose a way to investigate the region from dual conformal field theory by making use of AdS/CFT correspondence. We first construct gravastar geometry as an asymptotic anti-de Sitter spacetime, where the region inside the photon sphere is replaced by a horizon-less geometry. It is known that bulk-cone singularities in the retarded Green function in dual conformal field theory can encode the bulk null geodesics. We then compute numerically the retarded Green function from the bulk theory and observe bulk-cone singularities corresponding to null geodesics traveling into the interior region. In this way, we show that it is possible to examine the region inside the photon sphere from bulk-cone singularities of dual conformal field theory.

\end{abstract}
\vfill
\vskip 0.5 truecm

\setcounter{footnote}{0}
\renewcommand{\thefootnote}{\arabic{footnote}}
\end{titlepage}

\newpage

\hrule
\tableofcontents

\bigskip
\hrule
\bigskip

\section{Introduction}

AdS/CFT correspondence provides us a useful tool to examine the structure of bulk geometry from dual conformal field theory (CFT). 
In particular, we can learn the structure of Lorentzian spacetime from the singularities in the correlation functions of dual CFT. 
Let us consider the retarded Green functions of a CFT on $\mathbb{R}_t \times S^{d-1}$, where $\mathbb{R}_t$ denotes the time direction. The CFT may have two types of singularities. One of them is light-cone singularities (LC), which exist for any CFT when different boundary spacetime points become light-like separated. Another is bulk-cone singularities (BC)
such that boundary spacetime points become connected via bulk null geodesics.
The latter exists only in a strongly coupled CFT. In this note, we would like to read off the structure of bulk geometry from the bulk-cone singularities \cite{Hubeny:2006yu,Maldacena:2015iua,Dodelson:2020lal,Dodelson:2023nnr}. 
A non-trivial bulk structure is realized in the presence of black hole.
In the case of Schwarzschild black hole, a photon can travel around the photon sphere, which is located outside the black hole horizon. The photon sphere exists even in the case with asymptotic anti-de Sitter (AdS) space, and the bulk photon sphere can be captured by the bulk-cone singularities of dual CFT \cite{Hubeny:2006yu,Dodelson:2020lal,Dodelson:2023nnr}. See also, e.g.,  \cite{Hashimoto:2018okj,Hashimoto:2019jmw,Berenstein:2020vlp,Dodelson:2022eiz,Hashimoto:2023buz,Moitra:2024ixh,Caron-Huot:2025she}.

A black hole may be characterized by the existence of horizon. However, an outside observer may see only up to the location of photon sphere. 
For instance,
the observation by the Event Horizon Telescope (EHT) \cite{EventHorizonTelescope:2019dse} provided a picture around the photon sphere.
It was argued that it is difficult to distinguish a black hole from a possible alternative geometry, where the region inside the photon sphere is replaced by a horizon-less geometry \cite{Abramowicz:2002vt,Rosa:2024bqv}.
There are several candidates for black hole alternatives, such as ultra-compact boson stars, wormholes, and so on (see \cite{Cardoso:2019rvt} for a comprehensive review and references therein). As a concrete example, we shall focus on so-called  ``gravastar'' (gravitational-vacuum-star) \cite{Mazur:2001fv} (see also, e.g., \cite{Visser:2003ge,Pani:2009ss,Cardoso:2014sna}), where the geometry inside the horizon is replaced by de Sitter (dS) spacetime.
We first construct a gravastar geometry that asymptotically approaches the AdS spacetime.
We then investigate the corresponding retarded Green functions in the dual CFT, which are computed from the wave functions in the geometry by following \cite{Son:2002sd}.
In particular we numerically examine the singularity structure of the retarded Green functions and see how the difference of the geometries can be read off from them. 
Such black hole alternatives are expected to be unstable due to non-linear instabilities, see, e.g., \cite{Cunha:2022gde}. This may guarantee the ``uniqueness'' of black hole.
We hope that our setup allows us to examine this issue from the viewpoints of dual CFT in more detail.

The organization of this note is as follows.
In the next Section, we introduce the AdS-Schwarzschild black hole and study the null geodesics on the geometry. We then examine the wave equation for the scalar field on the geometry and explain how to read off the retarded Green function of dual CFT.
In Section \ref{sec:AdSgravastar}, we provide a generic method to construct AdS gravastar  and apply it to obtain the simplest model called ''thin-shell'' model.%
\footnote{
We restrict our attention to the specific case $\sigma=0$ of the thin-shell models, where $\sigma$ represents the surface energy density on the shell.} 
We then examine the null geodesics in the geometry and in particular point out that there exist null geodesic trajectories traveling the region inside the photon sphere. In Section \ref{sec:numerical}, we summarize our results of numerical computations on the retarded Green functions of dual CFT. We observe the bumps corresponding to both null geodesics traveling outside and inside the photon sphere. The observation of the bumps corresponding to the latter implies that the geometry is given by a geometry which is not a black hole. We also examine the case of gravastar without a photon sphere. There are bumps which may correspond to fictitious null geodesics reflected by the shell of the gravastar. Section \ref{sec:conclusion} is devoted to the conclusion and discussion. In Appendix \ref{app:AdSgravastar}, we present some technical details on the construction of AdS gravastar and its stability under small perturbations. We then provide concrete examples of null geodesics in the AdS gravastar, which appear in the main context. In Appendix \ref{app:wave}, we examine the relation between the null geodesics and the analysis based on wave equations. We further provide the explicit solution of wave equations in the case of pure dS spacetime.

\section{AdS-Schwarzschild black hole}
\label{sec:BH}

We begin by briefly reviewing useful results on the case with AdS-Schwarzschild black hole. The purposes of this Section are two fold. The first is to present results which will be compared with those obtained in the case of AdS gravastar. 
The second is to introduce the convention and method used for later analysis.

We start with the $(d+1)$-dimensional spacetime metric parameterized through warp factor $f(r)$ as:
\begin{align} \label{eq:metric}
    ds^2 = - f(r) dt^2 + f (r)^{-1} dr^2 + r^2 d \Omega_{d-1}^2 .
\end{align}
The metric of $(d-1)$-dimensional sphere is described by
\begin{align} \label{eq:round}
d\Omega_{d-1}^2 = d \theta^2 + \sin ^2 \theta d \Omega_{d-2}^2.
\end{align}
We will deal with the cases of the AdS-Schwarzschild black hole and AdS gravastar, and the metrics of these geometries can be put into the form of \eqref{eq:metric}.
For the AdS-Schwarzschild black hole, the function $f(r)$ is given by%
\footnote{
Here we set the AdS radius to be unity, i.e., $R_\text{AdS} = 1$. The AdS radius is related to the cosmological constant $\Lambda_\text{AdS}$ by $\Lambda_\text{AdS} = - d (d-1)/R_\text{AdS}^2$. 
Later we use the dS radius $R_\text{dS}$, which is related to the cosmological constant $\Lambda_\text{dS}$ by $\Lambda_\text{dS} = d (d-1)/R_\text{dS}^2$.}
\begin{align} \label{eq:AdSS-f}
        f(r) = r^2 + 1 - \frac{\mu}{r^{d-2}} .
\end{align}
The horizon is located at $r = r_h$, which satisfies
\begin{align} \label{eq:horizon}
    f(r_h) = r^2_h + 1 - \frac{\mu}{r^{d-2}_h} = 0  .
\end{align}
For AdS gravastar, we shall use the function $f(r)$ in \eqref{eq:gravastar-f}.

We would like to examine the structure of bulk geometry from the viewpoints of dual CFT. We put a CFT on $\mathbb{R}_t \times S^{d-1}$.
We examine the retarded Green function of scalar operator
\begin{align}
G_R (t , \theta) = i H(t) \langle [\mathcal{O} (t , \theta) ,  \mathcal{O} (0,0) ] \rangle ,
\end{align}
where $H(t)$ is the Heaviside step function.
The Green function may be computed from the propagator of bulk scalar field on AdS-Schwarzschild black hole. 
In order to see the behavior of the bulk propagator, we examine null geodesics and wave equations in this Section.
It is known that the retarded Green function has two types of singularities, see, e.g. \cite{Hubeny:2006yu}. One type is the ordinary light-cone singularity as 
\begin{align}
G_R (t , \theta) \propto  \frac{1}{(t - \theta)^\Delta}
\end{align}
near $t \sim \theta$.
Here $\Delta$ is the conformal dimension of CFT scalar operator, which is related to the mass $m$ of bulk scalar field as
\begin{align} \label{eq:Delta}
    \Delta = \frac{d}{2} + \sqrt{\frac{d^2}{4} + m^2} \, .
\end{align}
This type of singularity can be explained by the propagation of bulk scalar field along the AdS boundary. Another type is related to the null geodesics of scalar particle propagating inside the bulk geometry. In the case of pure AdS, the bulk-cone singularity coincides with the light-cone one, but they are different generically.

\subsection{Null geodesics}
\label{sec:null}

\begin{figure}[ht]
\centering
\includegraphics[width=8cm]{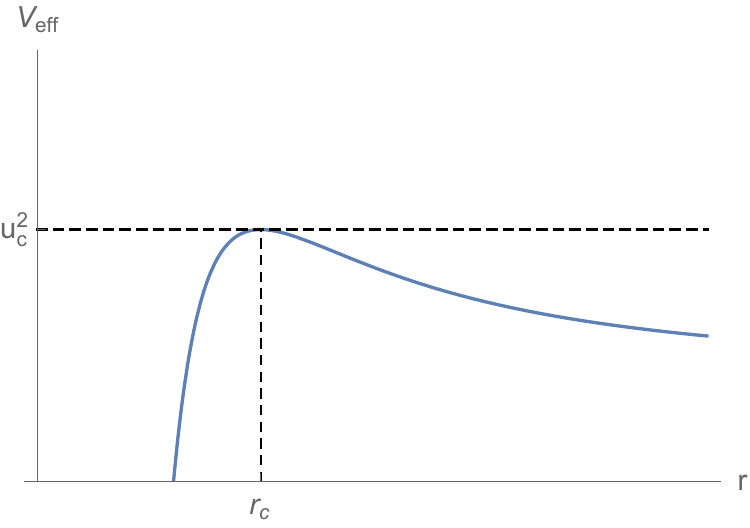}
\caption{The effective potential~\eqref{eq:geodpotential} for the AdS-Schwarzschild spacetime. The potential takes a maximal value $V(r_c) =u_c^2 $ at $r = r_c$.
}
\label{fig:adssch_potential}
\end{figure}

Let us consider the geometry with the metric of the form \eqref{eq:metric}.
In the following, we focus on the case with $d=3$,
where \eqref{eq:round} is given by where 
\begin{align}
d \Omega^2 \,(= d \Omega_2^2) = d\theta^2 + \sin^2 \theta d \phi^2 .
\end{align}
We introduce the geodesic tangent $k^\mu=dx^\mu/d\lambda$ with the affine parameter $\lambda$.
Its null condition can be written as
\begin{align}
\label{eq:geodesic-eq}
    g_{\mu\nu}k^\mu k^\nu
    =-f(r)\dot t^2+f(r)^{-1}\dot r^2+r^2\dot\theta^2
    =0 ,
\end{align}
where the dot means the derivative with respect to the affine parameter, i.e., $\dot A = d A/d \lambda$.
Furthermore, we have assumed that null geodesics are confined in the longitudinal plane, $\phi=0$ and $\phi=\pi$, without loss of generality.
Due to the existence of Killing vectors $\partial_t$ and $\partial_\theta$, we have two constants of motion, the energy $E$ and the angular momentum $L$, which are given by
\begin{align}
 E=-k_t=f(r)\dot{t} , \quad    L=k_\theta=r^2\dot\theta .
\end{align}
Substituting them into Eq.~\eqref{eq:geodesic-eq}, we obtain the radial equation of motion,
\begin{align}
\dot r^2+L^2f(r)r^{-2}=E^2.
\end{align}
Rescaling the affine parameter $\lambda\to\lambda/L$,  the equation reduces to
\begin{align}
\label{eq:geodpotential}
    \dot r^2+V_\mathrm{eff}(r)=u^2,\;\; V_\mathrm{eff}(r):= f(r)r^{-2},
\end{align}
where $u=E/L$ is the normalized energy for $L\neq0$.
This implies the trajectory of a null geodesic is characterized by the single parameter $u$.
The allowed region for $r$ is given by $V_\mathrm{eff}(r)<u^2$ and the turning point by $V_\mathrm{eff}(r)=u^2$ as in Fig.~\ref{fig:adssch_potential}.

The null geodesic equations are written in terms of $\lambda$ derivatives. Expressing in terms of $r$ derivatives, we have
\begin{align}
    \frac{d t}{d r} = \pm \frac{1}{f(r)} \frac{u}{\sqrt{u^2- V_\mathrm{eff}(r)}} , \quad 
    \frac{d \theta}{d r} = \pm \frac{1}{r^2} \frac{1}{\sqrt{u^2- V_\mathrm{eff}(r)}} .
\end{align}
Denoting the largest $r$ satisfying $V_\text{eff}(r) = u^2$ by $r_+$, the arrival time and angle of a bulk null geodesic from the boundary to boundary are expressed as
\begin{align} \label{eq:TTheta}
    T (u) = 2u \int_{r_+}^\infty \frac{dr}{f(r)} \frac{1}{\sqrt{u^2- V_\mathrm{eff}(r)}} , \quad
    \Theta (u) = 2  \int_{r_+}^\infty  \frac{dr}{r^2} \frac{1}{\sqrt{u^2- V_\mathrm{eff}(r)}} .
\end{align}

\begin{figure}[ht]
\centering
\includegraphics[width=5cm]{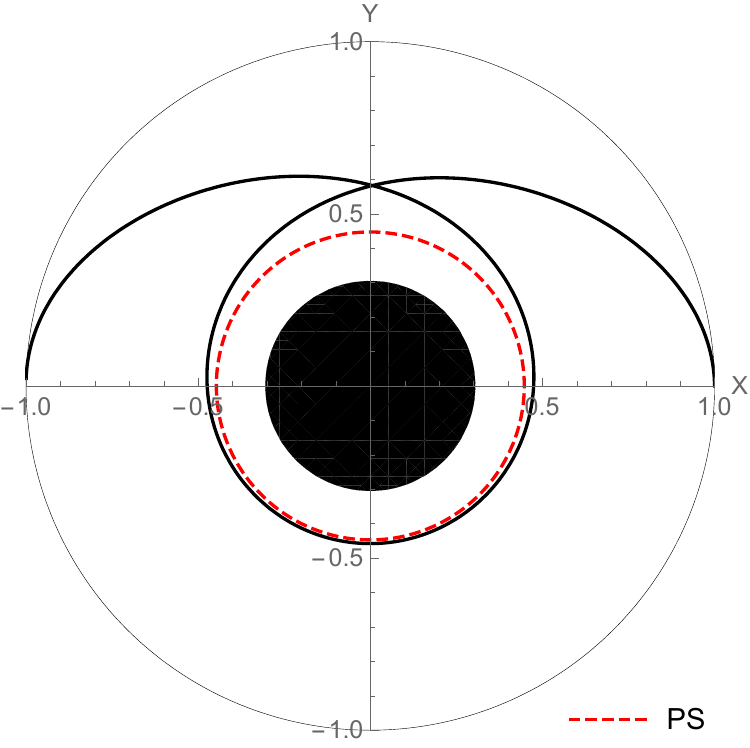}
\includegraphics[width=5cm]{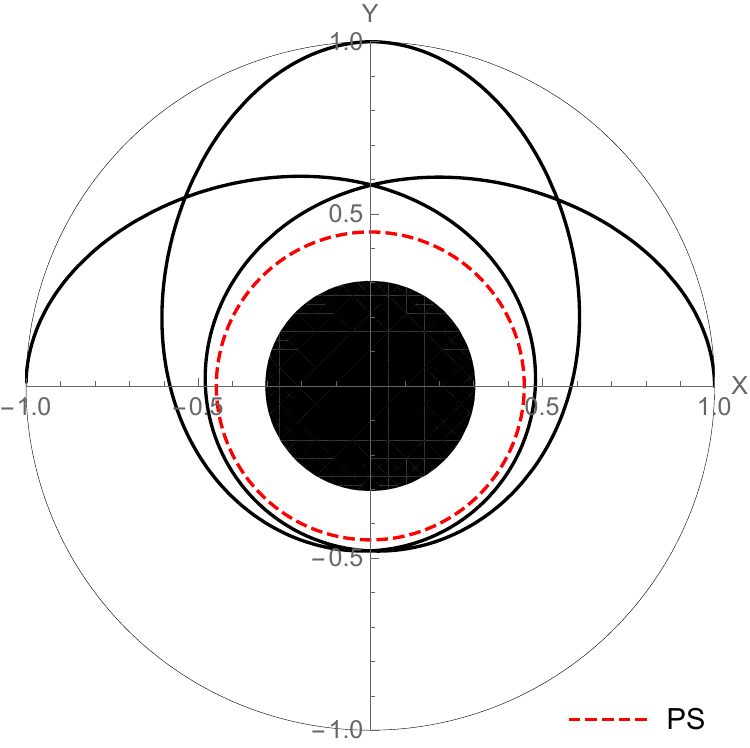}
    \caption{Left: A winding orbit around the photon sphere in the AdS-Schwarzschild spacetime.
    Right: An orbit bounced once by the conformal boundary.
    To draw the trajectories in a finite region, we adopted the coordinates $X=(r/\sqrt{1+r^2})\sin\theta$ and $Y=(r/\sqrt{1+r^2})\cos\theta$, where the AdS boundary is given by $X^2+Y^2=1$. 
    }
\label{fig:orbit-schwarz}
\end{figure}

Let us consider the case of AdS-Schwarzschild black hole, where the function $f(r)$ is given by \eqref{eq:AdSS-f}.
A circular orbit on the photon sphere is realized by the condition $\dot r=\ddot r=0$, which is equivalent to $V_\mathrm{eff}(r)=u^2$ and $\frac{d}{dr}V_\mathrm{eff}(r)=0$.
The condition determines the radius of photon sphere $r_c$ and the corresponding energy as
\begin{align} \label{eq:circular}
   \frac{d}{dr} [f(r)r^{-2}]|_{r_c}=0,\quad
    u_c^2=f(r_c)r_c^{-2}.
\end{align}
We obtain a unique solution to the equations as
\begin{align} \label{eq:photonsphere}
    r_c=\frac{3}{2}\mu>r_0,\quad
    u_c=\frac{\sqrt{4+27\mu^2}}{3\sqrt{3}\mu} .
\end{align}
Since $\frac{d^2}{dr^2}V_\mathrm{eff}(r) < 0$ at $r = r_c$, the circular orbit on the photon sphere is unstable.
This implies the existence of bulk null geodesics winding several times around the photon sphere as depicted in the left figure of Fig.~\ref{fig:orbit-schwarz}.

\begin{figure}
\centering
\includegraphics[height=6cm]{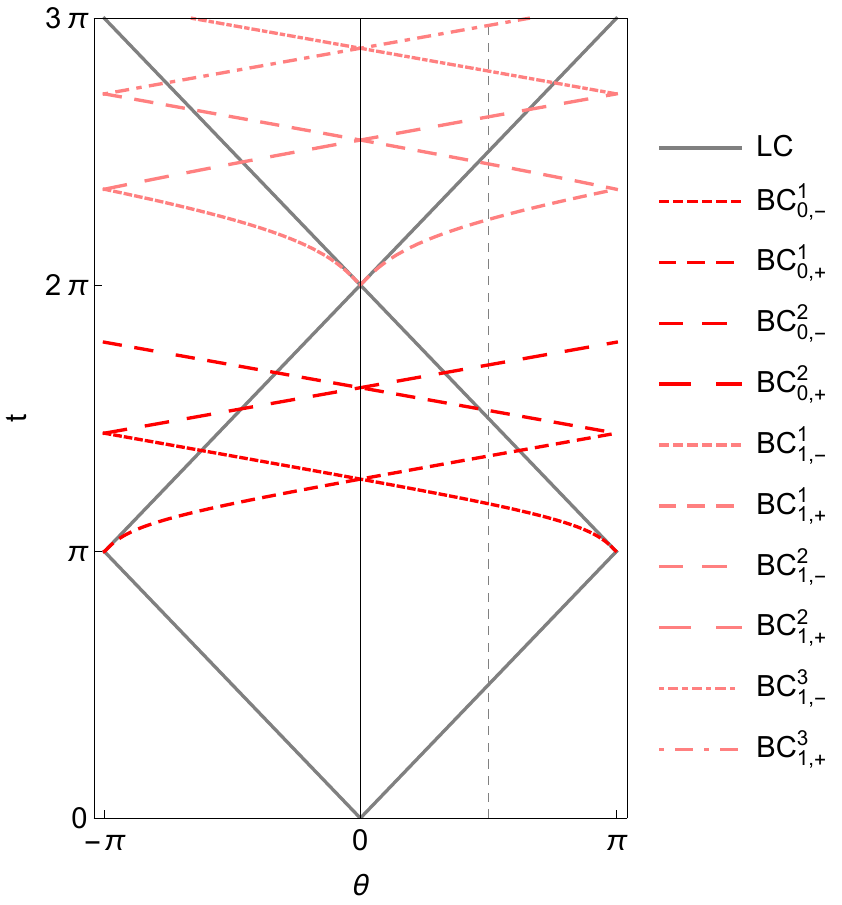}
\caption{The structure of bulk-cone singularities for the AdS-Schwarzschild black hole. The parameter is set as $\mu=1/15$.}
\label{fig:bcstructure_BH}
\end{figure}

In order to extract quantitative features of null geodesic, we use the arrival time and angle computed by
\begin{align} \label{eq:TTheta2}
    t=n T(u), \quad
    2\pi j \pm \theta=n \Theta(u),
\end{align}
where $T(u)$ and $\Theta (u)$ are defined in \eqref{eq:TTheta}.
The null geodesic is assumed to emerge at $(0,0)$ and reach $(t,\theta)$ on the boundary. 
The parameters $j$ and $n-1$ count the number of winding around the photon sphere and the number of bounce by the boundary, respectively.%
\footnote{See the right figure of Fig.~\ref{fig:orbit-schwarz} for an orbit bouncing at the boundary. To be precise, a null geodesic terminates at the conformal boundary with infinite value of the affine parameter. However, as the arrival time is finite, we can naturally extend the orbit as one reflected by the boundary.}
The sign in the right hand side represents the direction to round the photon sphere.
Each cone is labeled as $\mathrm{BC}_{n-1,\pm}^j$.
Fig.~\ref{fig:bcstructure_BH} shows the expected bulk-cone structure for the black hole case with $\mu=1/15$.%
\footnote{As in Fig.~\ref{fig:bcstructure_BH}, hereafter we extend the range of the polar angle from $0<\theta<\pi$ to $-\pi<\theta<\pi$ to cover the both $\phi=0$ and $\phi=\pi$ planes.
Any null geodesic can be regarded as an orbit confined in this longitudinal plane by rotation of the coordinates.
}

\subsection{Wave equations}
\label{sec:HGF}

We would like to compute the retarded Green function of scalar operator dual to a scalar field on the AdS-Schwarzschild black hole in a holographic way by following \cite{Son:2002sd}.
For this, we examine wave functions for the bulk scalar field. We consider the mode expansion of the scalar field as
\begin{align}
 \Phi (t , \theta, \phi ,r) = e^{- i \omega t} Y_{\ell m} (\theta , \phi) \frac{1}{r} \psi_{\omega \ell} . 
\end{align}
Here the spherical harmonics on $S^{2}$ are denoted by $Y_{\ell m}(\theta ,\phi)$, where $\ell$ and $m$ are the orbital angular momentum and the azimuthal angular momentum, respectively. The frequency for the Fourier transformation is represented by $\omega$.
The wave equation for the radial part can be put into the form%
\footnote{The potential $V(z)$ is introduced here by putting the radial wave equation into the form of Schr\"{o}dinger equation. On the other hand, the potential $V_\text{eff}(r)$ in \eqref{eq:geodpotential} was given by rewriting the radial null equation. These two potentials are different but closely related. In fact, we can show that the potential $V(z)$ in the eikonal limit reduces to $V_\text{eff}(r)$, see, e.g., Appendix \ref{app:Eikonal}.}
\begin{align}
\label{eq:zwaveeqp}
    (- \partial_z^2 + V(z) - \omega^2) \psi_{\omega \ell} (z) = 0  .
\end{align}
The tortoise radial coordinate $z$ is introduced via
\begin{align} \label{eq:zcoordinate}
    z =  \int^\infty_r \frac{dr'}{f(r')} 
\end{align}
and the potential $V(z)$ is%
\footnote{Here the function $V(z)$ of $z$ is defined by the right hand side of \eqref{eq:potential} in terms of $r$ via the transformation \eqref{eq:zcoordinate}.}
\begin{align} \label{eq:potential}
    V(z)=f(r)\left[\frac{(\ell+\alpha)^2-\frac{1}{4}}{r^2}+ \nu^2 - (\alpha + 1)^2 +\left(\alpha^2 - \frac14\right)\frac{f(r)-1}{r^2}+\left(\alpha + \frac12\right)\frac{\frac{d}{dr}f(r)}{r}\right] ,
\end{align}
where $d =3$ and $\alpha = (d-2)/2 = 1/2$. Moreover, we set 
\begin{align} \label{eq:nu}
\nu = \Delta - \frac{d}{2} = \Delta - \frac{3}{2}
\end{align}
with
\eqref{eq:Delta}.
For the case with \eqref{eq:AdSS-f} and $d=3$, the potential $V(z)$ becomes
\begin{align}
    V(z) = f(r) \left[ \frac{\ell (\ell + 1)}{r^2} + \nu^2 - \frac14 + \frac{ \mu}{ r^3}\right]  .
\end{align}
In terms of the new coordinate $z$ in \eqref{eq:zcoordinate}, the black hole horizon and the AdS boundary are located at $z \to \infty$ and $z \to 0$, respectively. At the black hole horizon $(z \to \infty)$, we assign the ingoing boundary condition
\begin{align}
    \psi_{\omega \ell} \sim e^{i \omega z}  . 
\end{align}
At the AdS boundary $(z \to 0)$, the scalar field behaves as
\begin{align}\label{eq:AdSasym}
    \psi_{\omega \ell} \sim \mathcal{A} (\omega , \ell) z^{\frac12 - \nu} + \mathcal{B} (\omega , \ell) z^{\frac12 + \nu}  .
\end{align}
The retarded Green function is then obtained as
\begin{align} \label{eq:retarded}
    G_R(\omega , \ell) = \frac{\mathcal{B} (\omega , \ell)}{\mathcal{A} (\omega , \ell)}  .
\end{align}
The poles of $G_R(\omega , \ell) $ are realized by $\mathcal{A} (\omega , \ell) = 0$ and the condition in the wave functions \eqref{eq:AdSasym} coincides with the one for quasi-normal modes. In \cite{Dodelson:2023nnr}, they computed the Green function from the semi-classical analysis of wave functions. The result coincides with the one obtained by numerical analysis.
We will explain the numerical analysis in more details in Section \ref{sec:numerical}.

\section{AdS gravastar}
\label{sec:AdSgravastar}

Inspired by the work of \cite{Mazur:2001fv}, the authors of \cite{Visser:2003ge} constructed a simpler version of gravastar which has three layers. For $r > r_0$, the geometry is given by Schwarzschild black hole. At $r=r_0$, we put a thin shell. For $r < r_0$, the geometry is given by a dS universe with $\rho = - p$. 
In Section \ref{sec:construction}, we briefly explain the construction of a simple model of AdS gravastar, where the region with $r > r_0$ is replaced by AdS-Schwarzschild black hole.
In Section \ref{sec:geodesic-potential}, we will examine the null geodesics in the geometry of AdS gravastar.
We relegate most of the computation details into Appendix \ref{app:AdSgravastar}.

\subsection{Construction}
\label{sec:construction}

Following \cite{Visser:2003ge}, we construct a three-layered geometry. We put a shell at $r = r_0$. The metrics for the exterior region with $r > r_0$ and the interior region $r < r_0$ are written as 
\begin{align} \label{eq:ansatz}
    ds^2_\pm=-f_\pm(r_\pm)dt_\pm^2+g_\pm(r_\pm)dr_\pm^2+r_\pm^2d\Omega^2.
\end{align}
The subscripts $+$ and $-$ indicate that the metrics and coordinates are those for the exterior and interior regions, respectively.
We assign the junction conditions at $r=r_0$. 
We first require that the induced metrics $h_\pm$,
\begin{align} \label{eq:induced}
    h_\pm=-f_\pm(r_0) dt_\pm^2+r_0^2d\Omega^2,
\end{align}
to be continuous, i.e., $h_+=h_-\equiv h$.
The condition gives the relation between the time coordinates,
\begin{align} \label{eq:cond_ind}
    \frac{dt_+^2}{dt_-^2}=\frac{f_-(r_0)}{f_+(r_0)}.
\end{align}
We then require that
\begin{align} \label{eq:cond_energy}
    [[\chi]]-[[\text{tr} \chi]]h=-8\pi S.
\end{align}
The bracket $[[A]]=A_+-A_-$ denotes the jump of a quantity $A$ across the joint boundary,
while the extrinsic curvature and its trace of the time-like hypersurface are:
\begin{align} \label{eq:extrinsic}
\begin{aligned}
    \chi_\pm=
    &g_\pm^{-1/2}(r_0)\left(-\frac{1}{2}f_\pm'(r_0)dt_\pm^2+r_0d\Omega^2\right),\\
    \text{tr}  \chi_\pm&=g_\pm^{-1/2}(r_0)\left(\frac{1}{2}\frac{f_\pm'(r_0)}{f_\pm(r_0)}+\frac{2}{r_0}\right) .
\end{aligned}
\end{align}
With the surface energy density $\sigma$ and surface tension $\bar p$, the surface stress energy tensor takes the form, 
\begin{align}
S=\sigma d\tau^2+\bar p r_0^2d\Omega^2,
\end{align}
due to the time independence and spherical symmetry.
Here $\tau$ is the proper time of the shell satisfying $d\tau\propto dt_\pm$.

We are interested in the case where the exterior region is given by the AdS-Schwarzschild black hole and the interior region is described by dS spacetime.
In order to realize these geometries, we set the metric coefficients as
\begin{align} \label{eq:fpfm}
\begin{aligned}
    f_+(r)&=g_+^{-1}(r)=1-\frac{\mu}{r}+\frac{\Lambda_+}{3} r^2,\\
    f_-(r)&=g_-^{-1}(r)=1-\frac{\Lambda_-}{3} r^2,
\end{aligned}
\end{align}
where we have defined both $\Lambda_\pm>0$ as positive constants.
Additionally, we need to specify the equation of state (EOS) of the shell, $\bar p=\bar p(\sigma)$, to construct the AdS gravastar.

In order for the analysis of the scalar field propagation to be tractable, we consider the simplest geometry among the ones constructed above. We set the surface energy density to vanish, i.e., $\sigma=0$. We call this setup  ``thin-shell'' model. See \cite{Pani:2009ss,Cardoso:2014sna} for the case with asymptotic flat spacetime and Appendix \ref{app:thinshell} for more detailed analysis. 
Our simplest model is summarized from the physical point of view as follows.
Let us introduce new parameters; the background AdS cosmological constant $\Lambda:=-\Lambda_+<0$, the energy density of the central condensate $\rho=(\Lambda_++\Lambda_-)/8\pi=\mathrm{const}.$, and its pressure $p=-\rho$.
The energy momentum tensor is expressed as
\begin{align}
    T_{\mu\nu}=
    \left\{\begin{array}{ll}
    -(\Lambda/8\pi) g_{\mu\nu} & (r>r_0),  \\
    \rho u_\mu u_\nu +p(g_{\mu\nu}+u_\mu u_\nu) -(\Lambda/8\pi) g_{\mu\nu} & (r<r_0), \\
    \delta(l) S_{\mu\nu} & (r=r_0),
    \end{array}\right.
\end{align}
where $l$ is a proper radial coordinate such that $dl$ coincides with the unit vector normal to the shell and $l=0$ is located on the shell.
The metric can be put into the form \eqref{eq:metric}, i.e.,
\begin{align}
    ds^2=-f(r)dt^2+f^{-1}(r)dr^2+r^2d\Omega^2.
\end{align}
The metric function $f(r)$ is given by
\begin{align}
\label{eq:gravastar-f}
    f(r)=\left\{\begin{array}{ll}
    1- \mu / r-(\Lambda /3 ) r^2 & (r\ge r_0), \\
    1-\left(8\pi \rho +( \Lambda / 3 )\right) r^2 & (r<r_0).
    \end{array}\right.
\end{align}
The function should be continuous at $r=r_0$, which leads to
\begin{align}
\label{eq:r0-mu-rho}
    r_0^3=\frac{3\mu}{8\pi \rho}.
\end{align}

For a horizon-less geometry, we have to assign $r_0 > r_h$. Here $r_h$ is the radius of horizon of the original AdS-Schwarzschild black hole satisfying \eqref{eq:horizon}. The condition leads to
\begin{align}
\label{eq:horizon-less-cond}
    f_+(r_0)>0.
\end{align}
Let us consider the gravastar which has a photon sphere. In this case, we say that the gravastar is small. The condition is given by
\begin{align}
\label{eq:ps-cond}
    r_0<r_c =\frac{3}{2}\mu,
\end{align}
where the photon sphere radius $r_c$ is given in \eqref{eq:photonsphere}.
If $r_0>r_c$, we say that the gravastar is large.
The conditions \eqref{eq:horizon-less-cond} and \eqref{eq:ps-cond} can be satisfied when $r_0$ satisfies (see \eqref{eq:r0-condition})
\begin{align}
    \frac{1}{\sqrt{12\pi \rho}}
    <r_0
    <\left(8\pi \rho+\frac{\Lambda}{3}\right)^{-1/2}.
\end{align}

To summarize, we have two independent parameters, the ADM mass $M=\mu/2$
and the gravastar radius $r_0$, for our model in addition to the AdS radius $R_\mathrm{AdS}=1$.
The energy density $\rho$ in the dS region is determined from these parameters through Eq.~\eqref{eq:r0-mu-rho}, which is also rewritten as the interior cosmological constant $\Lambda_-$ or the dS radius $R_\mathrm{dS}$.

Let us make several comments on the thin-shell model.
This model has an advantage that the time and radial coordinates defined in each region are smooth across the shell, i.e., $dt_+/dt_-=1$ and $dr_+/dr_-=1$.
This allows us to adopt the coordinates as the same coordinate system over the entire spacetime. 
This level of smoothness of the coordinate system guarantees the continuity of coordinate components of smooth vector fields.
Furthermore,
as analyzed in Appendix~\ref{sec:shell-dynamics}, the assumption $\sigma=0$ makes the shell non-dynamical.
Only a static configuration is allowed for the spherical shell in a $\Lambda$-vacuum spacetime.
Even for the case of $\sigma\neq0$, the geometry could have allowed parameter region of
 dynamical stability against radial perturbation, see, e.g., \cite{Visser:2003ge} for the asymptotic flat case.

\subsection{Null geodesics}
\label{sec:geodesic-potential}

\begin{figure}
\centering
\includegraphics[width=8cm]{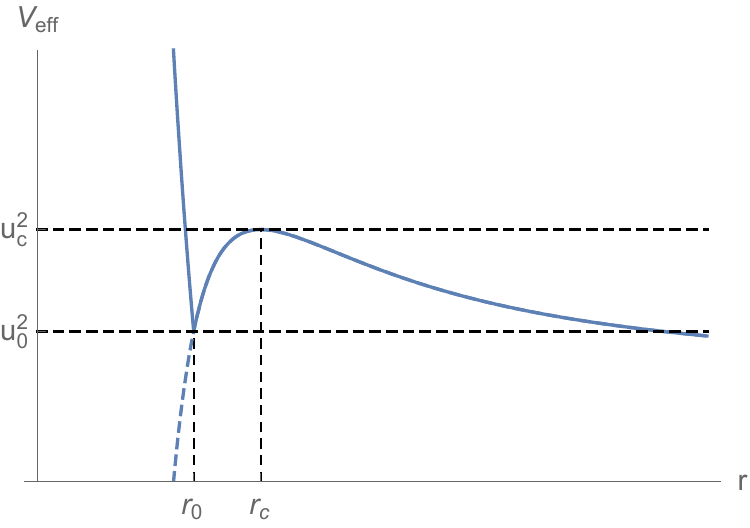}
\caption{
The effective potential~\eqref{eq:potential_gravastar} for the AdS gravastar. The dashed curve represents the AdS Schwarzschild case.
}
\label{fig:potential}
\end{figure}

As explained in Section \ref{sec:null}, the radial part of null geodesic equation can be put into the form  (see \eqref{eq:geodpotential})
\begin{align}
\label{eq:potential_gravastar}
\dot r^2 + V_\mathrm{eff}(r) = u^2 ,\quad V_\mathrm{eff}(r)=f(r)r^{-2} .
\end{align}
In the case of AdS gravastar, the function $f(r)$ is given by Eq.~\eqref{eq:gravastar-f}.
The typical shape of the potential $V_\mathrm{eff}(r)$ is shown in Fig.~\ref{fig:potential}.
The allowed region for $r$ is given by $u^2>V_\mathrm{eff}(r)$ and the turning point $r_+$ is realized by $V_\mathrm{eff}(r_+)=u^2$.
The photon sphere may exist outside the shell with $r > r_0$, where the geometry is realized by AdS-Schwarzschild black hole. Thus the condition \eqref{eq:circular} for the photon sphere leads to the same values as in \eqref{eq:photonsphere}.

In the gravastar geometry, the region inside the horizon is replaced by dS spacetime. Due to the centrifugal force, the height of potential goes to the positive infinity at $r \to 0$. This implies the existence of a minimum of the potential at $r=r_0<r_c$, see Fig.~\ref{fig:potential}.
A photon orbit can be trapped around this minimum, which is repetitively reflected by the potential barriers on the both sides.

\begin{figure}
\centering
\includegraphics[width=5cm]{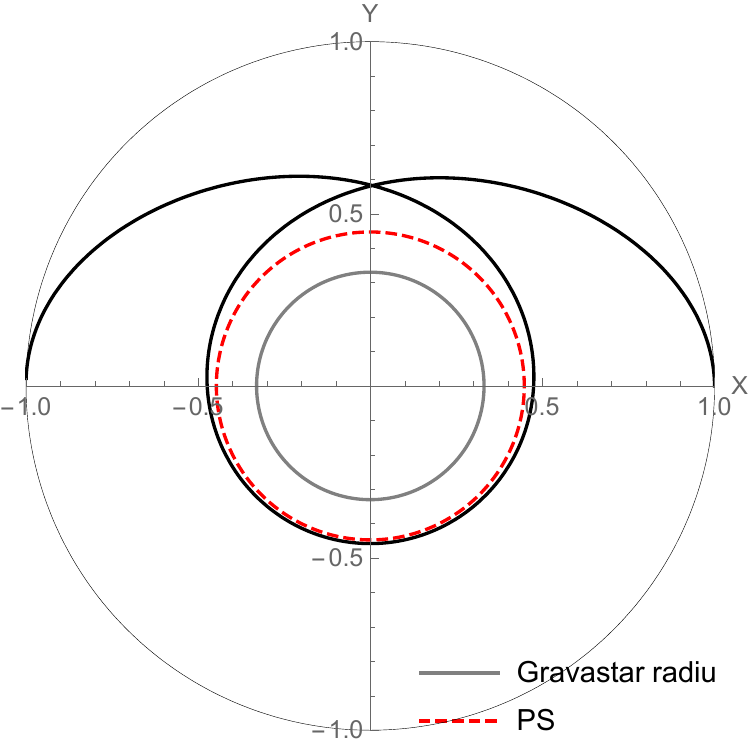}
\includegraphics[width=5cm]{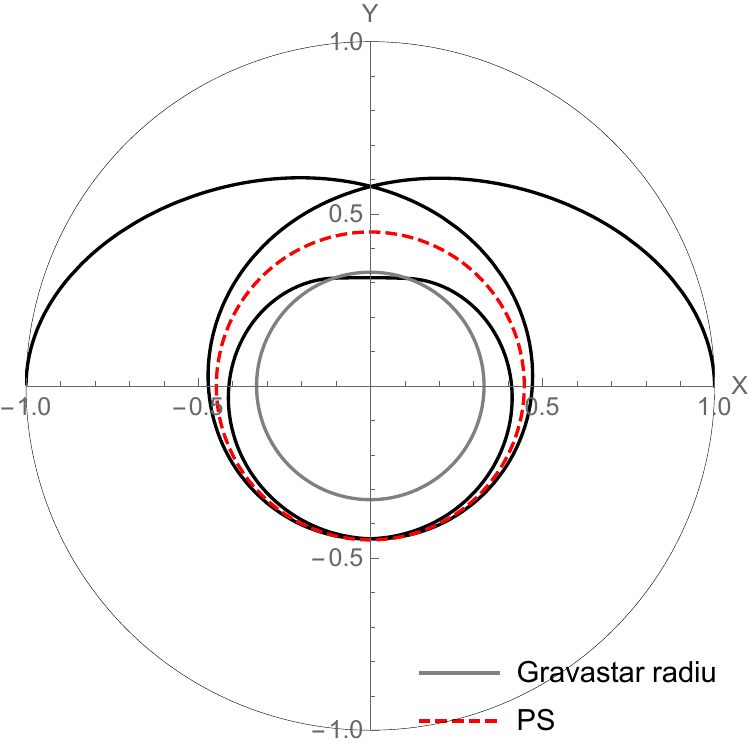}
    \caption{
    Outer (Left) and inner (Right) winding orbits.
    }
\label{fig:outer-winding-orbit}
\end{figure}

There are two kinds of characteristic orbits.
The first one is an orbit winding outside the photon sphere (see Fig.~\ref{fig:outer-winding-orbit}, Left).
It has subcritical energy, i.e. $u<u_c$,
and is reflected to infinity by the photon sphere potential.
The other one is an orbit winding both outside and inside the photon sphere (see Fig.~\ref{fig:outer-winding-orbit}, Right).
It has supercritical energy $u>u_c$.
This orbit passes through the photon sphere after winding several times, and due to the potential barrier inside the gravastar, it eventually escape to infinity.
This is characteristic of a horizon-less ultra-compact object.

\begin{figure}
\centering
\includegraphics[height=6cm]{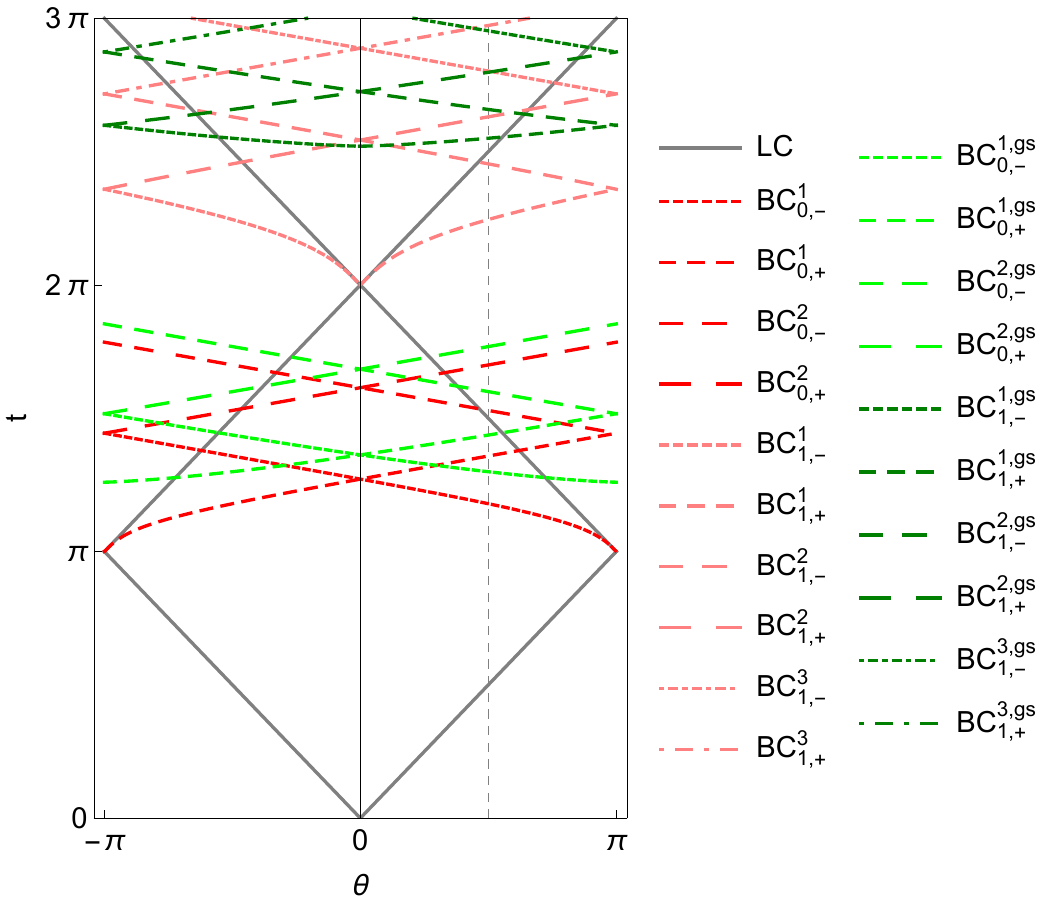}
\caption{The structure of bulk-cone singularities  for a small gravastar case. The parameters are set as $\mu=1/15$ and $r_0=0.07$.}
\label{fig:bcstructure_r0_007}
\end{figure}

As explained in Section \ref{sec:null}, we can quantify the null geodesics by the arrival time and angle computed with \eqref{eq:TTheta2}. See Fig.~\ref{fig:bcstructure_BH} for the case of AdS-Schwarzschild black hole.
The structure of bulk-cones 
for a small gravastar with $\mu=1/15$ and radius $r_0=0.07$ is shown in  Fig.~\ref{fig:bcstructure_r0_007}.
The green curves and the ones with superscript ``gs'' represent the null geodesics passing through the gravastar.
There are the first and second series of curves corresponding to the cases $n=1$ and $n=2$, respectively, for each type of bulk-cones.
Bulk-cones with higher $j$ are omitted to keep the figure simple.
The first and second series of null geodesic trajectories are obtained as in Figs.~\ref{fig:orbits-1stseries_r0_007} and \ref{fig:orbits-2ndseries_r0_007}, respectively.
The energy levels with respect to the effective potential are described in Fig.~\ref{fig:energylevels-1stseries_r0_007}.

\begin{figure}
\centering
\includegraphics[height=6cm]{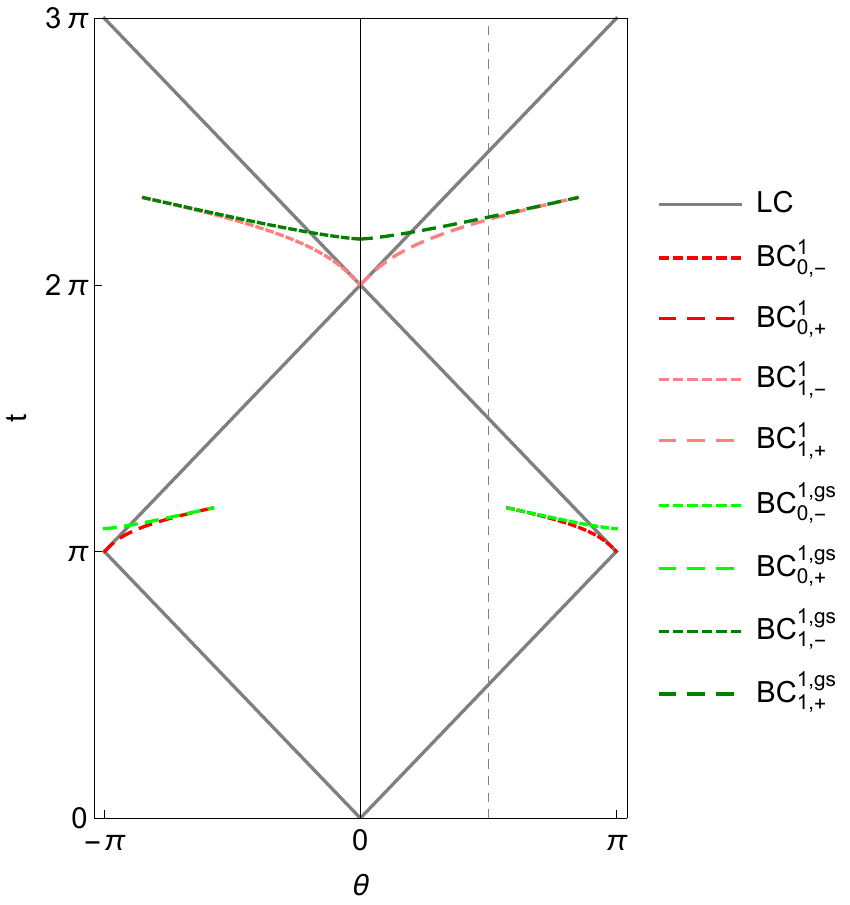}
\includegraphics[height=6cm]{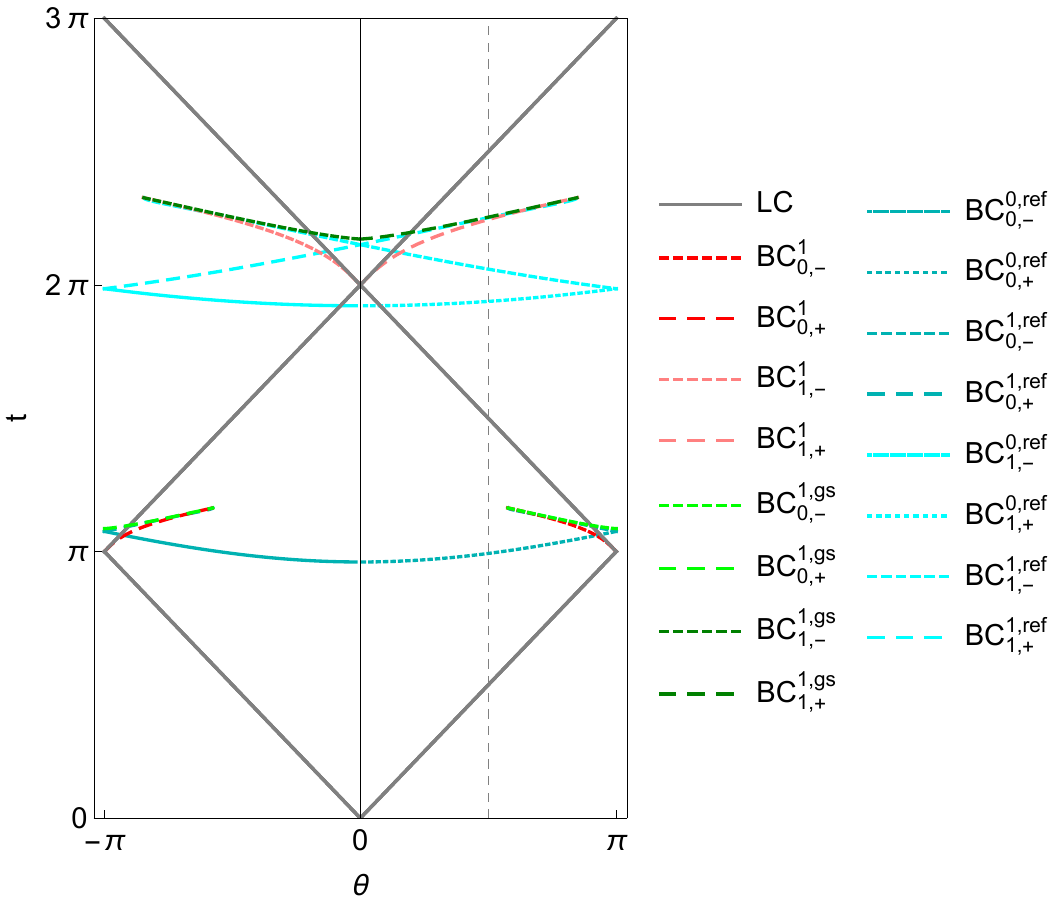}
\caption{The structure of bulk-cone singularities  for a large gravastar case. The parameters are set as $\mu=1/15$ and $r_0=0.17$. The right figure includes fictitious null geodesics that are reflected by the gravastar surface.}
\label{fig:bcstructure_r0_017}
\end{figure}

Setting $r_0 > r_c$, we can construct a large gravastar without a photon sphere. Because of the absence of a photon sphere, we expect that the structure of bulk-cones is much different from that for a small gravastar. The structure of bulk-cone singularities  for a large gravastar with $\mu=1/15$ and radius $r_0=0.17$ is shown in  Fig.~\ref{fig:bcstructure_r0_017}.
In this case, bulk-cones of the first series do not appear at $\theta=\pi/2$ for example.
As we shall observe later in the numerical analysis of retarded Green functions, the wave functions of scalar field could be partially reflected at the junction points. 
The cyan curves and the ones with subscript ``ref'' show fictitious null geodesics that are reflected by the shell.
The first series of null geodesic trajectories and the energy levels with respective to the effective potential are obtained as in Figs.~\ref{fig:orbits-1stseries_r0_017}.
Similarly the second series of null geodesic trajectories is obtained as in Figs.~\ref{fig:orbits-2ndseries_r0_017}
and the energy levels with respective to the effective potential are described in Fig.~\ref{fig:energylevels-2ndseries_r0_017}.

\section{Numerical computation of Green functions}
\label{sec:numerical}

In Section \ref{sec:HGF}, we explained how to compute the retarded Green function from dual bulk theory in the case of AdS-Schwarzschild black hole. In this Section, we apply this method to numerically compute the retarded Green functions in the cases of AdS gravastar. 
We solve the wave equation \eqref{eq:zwaveeqp}, where the potential is given as (see \eqref{eq:potential})
\begin{align} \label{eq:gspotential}
    V(z)=f(r)\left[\frac{\ell (\ell +1)}{r^2}+\nu^2-\frac{9}{4}+\frac{\frac{d}{dr}f(r)}{r}\right] .
\end{align}
Here we also set
the function $f(r)$ as in \eqref{eq:gravastar-f}.

The retarded Green function in the frequency domain is given by Eq.~\eqref{eq:retarded} with the asymptotic solution~\eqref{eq:AdSasym}.
We express the asymptotic solution as
\begin{align}
    \psi_{\omega \ell}(z)\sim \mathcal{A}(\omega,\ell)\psi_\mathcal{A}(z)+ \mathcal{B}(\omega,\ell)\psi_\mathcal{B}(z),
\end{align}
where
\begin{align}
\label{eq:expansion}
    \psi_\mathcal{A}(z)=z^{\frac12-\nu}\left(1+\sum_{n=2}C^\mathcal{A}_nz^n\right),
     \quad \psi_\mathcal{B}(z)=z^{\frac12+\nu}\left(1+\sum_{n=2}C^\mathcal{B}_nz^n\right).
\end{align}
For the numerical computation, we set the mass of scalar field to vanish, i.e., $m=0$. Eqs.~\eqref{eq:Delta} and \eqref{eq:nu} imply $\nu = 3/2$.
With the value of $\nu$, the asymptotic solution is given by%
\footnote{
In this case, the coefficient $C_3^\mathcal{A}$ of $\psi_\mathcal{A}$ is not determined as it is degenerated with the degree of freedom of $\psi_\mathcal{B}$.
We have set $C_3^\mathcal{A}=0$ so that the coefficient $\mathcal{A}C_3^\mathcal{A}+\mathcal{B}$ of the $z^3$ term becomes $\mathcal{B}$.} 
\begin{align}
    \psi_{\omega \ell}(z)=z^{\frac12-\nu}\left(\mathcal{A}+\mathcal{A}C^\mathcal{A}_2z^2+\mathcal{B}z^3+\mathcal{O}(z^4)\right).
\end{align}
Given a numerical solution of $\psi_{\omega \ell}(z)$ with suitable boundary condition at a horizon or center, we can read off the coefficient as
\begin{align}    \mathcal{A}=\lim_{z\to0}\left(z^{-\frac12+\nu}\psi_{\omega \ell}(z)\right),\quad
\mathcal{B}=\lim_{z\to0}\frac16\frac{d^3}{dz^3}\left(z^{-\frac12+\nu}\psi_{\omega \ell}(z)\right).
\end{align}
From them, we obtain $G_R(\omega,\ell)=\mathcal{B}/\mathcal{A}$ as in \eqref{eq:retarded}.

The time-domain Green's function is obtained by $\omega$-integral and $\ell$-sum as
\begin{align}
    G_R(t,\theta)=\int_{-\infty + i \delta}^{+\infty + i \delta}d\omega e^{-i\omega t}\sum_{\ell=0}^{\infty}Y_{\ell 0}(\theta)G_R(\omega,\ell)e^{-\frac{\mathrm{Re}(\omega)^2}{\omega_c^2}-\frac{\ell^2}{\ell_c^2}}
\end{align}
with $\delta > 0$.
As there might be poles on or slightly below the real axis, we take the contour slightly above the real axis.
The Gaussian factor is the smoothing factor with the cut-off parameters $\omega_c$ and $\ell_c$ so that the singularities reduce to finite bumps.
Using the fact $G_R(-\omega,\ell)=G_R(\omega,\ell)^*$, we have
\begin{align}
    G_R(t,\theta)=\int_{0+i\delta}^{\infty+i\delta}d\omega \sum_{\ell=0}^{\infty}Y_{\ell0}(\theta)\left[G_R(\omega,\ell)e^{-i\omega t}+G_R(\omega,\ell)^*e^{i\omega^* t}\right]e^{-\frac{\mathrm{Re}(\omega)^2}{\omega_c^2}-\frac{\ell^2}{\ell_c^2}}.
\end{align}
As $G_R(\omega,\ell)$ is analytic in upper half $\omega$-plane, any positive deformation parameter $\delta \ (>0)$ is allowed.
We will perform the summation over $\ell$ and the integration over $\omega$ up to the sufficiently large values $\ell_\mathrm{max}\gg \ell_c$ and $\omega_\mathrm{max}\gg\omega_c$, respectively.

\begin{figure}
\centering
\includegraphics[width=16cm]{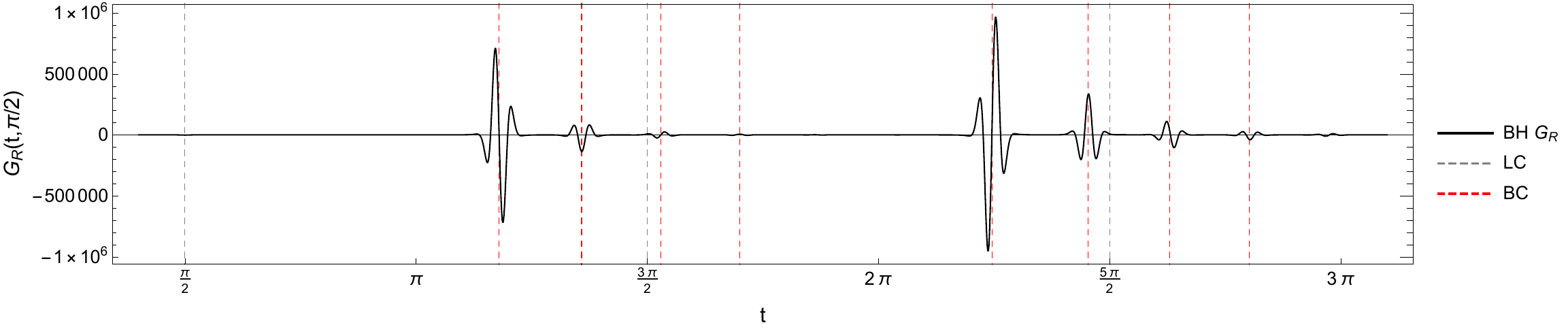}
\caption{The Green function $G_R(t,\pi/2)$ for the black hole case. The parameters are set as $R_\mathrm{AdS}=1$, $\nu=3/2$, $\mu=1/15$, $\omega_c=\ell_c=40$, and $\omega_\mathrm{max}=\ell_\mathrm{max}=150$.
The arbitrary parameter in the integration is taken to be $\delta=0.5$.
The vertical dashed lines are for LC singularities (gray) and BC singularities expected from the geodesic analysis (red).
The LC singularities have small amplitude and are difficult to see at this scale.
The BC singularities are labeled as $\mathrm{BC}^1_{0,-}$, $\mathrm{BC}^1_{0,+}$, $\mathrm{BC}^2_{0,-}$, $\mathrm{BC}^2_{0,+}$, $\mathrm{BC}^1_{1,+}$, $\mathrm{BC}^2_{1,-}$, $\mathrm{BC}^2_{1,+}$, and $\mathrm{BC}^3_{1,-}$ in time order.
}
\label{fig:GRt_BH}
\end{figure}

Before going into the gravastar cases,
we show the numerical result of $G_R(t,\pi/2)$ for four-dimensional AdS-Schwarzschild black hole in Fig.~\ref{fig:GRt_BH}.
The bumps represent the smeared singularities.
The gray dashed lines represent the light-cone (LC) singularities.
Although there are bumps on them, the amplitudes are so small that we cannot see them in the scale of the figure.
The red dashed lines represent the bulk-cone (BC) singularities that are estimated from the arrival times of null geodesics from $\theta=0$ to $\pi/2$ on the boundary, see \eqref{eq:TTheta2}.
The first series of the bumps appears in $\pi< t< 2\pi$.
They are associated with the null geodesic trajectories that are not bounced by the boundary.
The second series appears in $2\pi< t< 3\pi$.
They are associated with the trajectories bounced by the boundary once.
The result agrees with that of Fig.~11 of \cite{Dodelson:2023nnr} up to the overall scaling.

\subsection{Small Gravastar}
\label{sec:smallgravastar}

We examine the retarded Green functions from the wave functions of scalar field in small gravastar geometry. We impose the regularity condition of the wave function at the center.
This is equivalent to the condition that the growing mode toward the center vanishes.
Since the interior region of our gravastar solution is given by the dS spacetime, we have the exact solution for the region near the center. We summarize relevant results in Appendix \ref{app:dS}. For the interior region, the wave function satisfying the regularity condition is obtained as
\begin{align} \label{eq:psi}
    \psi&=\left(\frac{r}{R_\mathrm{dS}}\right)^{\frac{d-1}{2}+\ell}\Bigl[
    A(1-(r/R_\mathrm{dS})^2)^{\frac{i\omega R_\mathrm{dS}}{2}}{}_2F_1(1+a-c,1+b-c;2-c;1-(r/R_\mathrm{dS})^2)\nonumber\\
    &\qquad +B(1-(r/R_\mathrm{dS})^2)^{\frac{-i\omega R_\mathrm{dS}}{2}}{}_2F_1(a,b;c;1-(r/R_\mathrm{dS})^2)
    \Bigr]\nonumber\\
    &=:\psi_\mathrm{dS},
\end{align}
where we set%
\footnote{We can choose any $A$ and $B$ that satisfy Eq.~\eqref{eq:regularity-BA}. Here we have taken specific normalization of the wave function. This does not affect the Green function~\eqref{eq:retarded}.}
\begin{align} \label{eq:psicoeff}
\begin{aligned}
    a&=\frac34+\frac \ell2-\frac14 \sqrt{9-4m^2R_\mathrm{dS}^2}-\frac{i\omega R_\mathrm{dS}}{2},\\
    b&=\frac34+\frac \ell2+\frac14 \sqrt{9-4m^2R_\mathrm{dS}^2}-\frac{i\omega R_\mathrm{dS}}{2},\\
    c&=1-i\omega R_\mathrm{dS},\\
    A&=\Gamma(1+a-c)\Gamma(1+b-c)\Gamma(c),\\
    B&=-\Gamma(a)\Gamma(b)\Gamma(2-c).
\end{aligned}
\end{align}
At the shell radius $r_0$, the boundary condition is translated to 
\begin{align}
&\psi(z_0)=\psi_\mathrm{dS}(z_0) , \\
&\left .\frac{d\psi}{dz} \right|_{z=z_0} =
\left .\frac{d\psi_\mathrm{dS}}{dz} \right|_{z=z_0} = \left. \frac{dr}{dz} \frac{d\psi_\mathrm{dS}}{dr} \right|_{z = z_0}=- \left. f(r_0) \frac{d\psi_\mathrm{dS}}{dr} \right|_{z = z_0}  .
\end{align}
Here $z_0$ is the shell radius in the tortoise coordinate.
The numerical solution of $\psi(z)$ in the exterior region $0<z<z_0$ is obtained by integrating the wave equation with this initial condition at $z=z_0$.

\begin{figure}
\centering
\includegraphics[width=16cm]{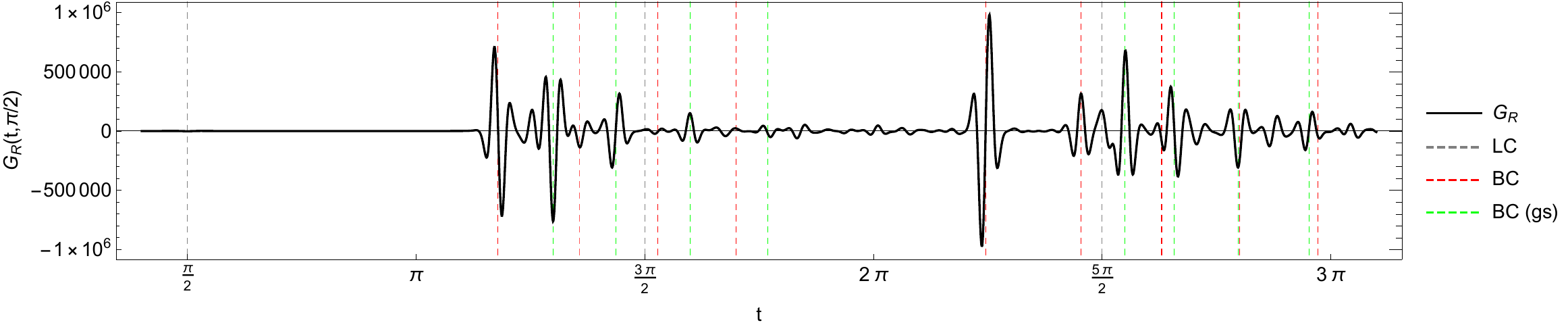}
\caption{The Green function $G_R(t,\pi/2)$ for the small gravastar. The vertical dashed lines indicate the arrival times of null geodesics.
Red lines describe the orbits reflected by the photon sphere potential. The green lines are for the orbits transmitting into the interior region described by the dS spacetime.}
\label{fig:GRt_GV_r0_007}
\end{figure}

\begin{figure}
\centering
\includegraphics[width=7.9cm]{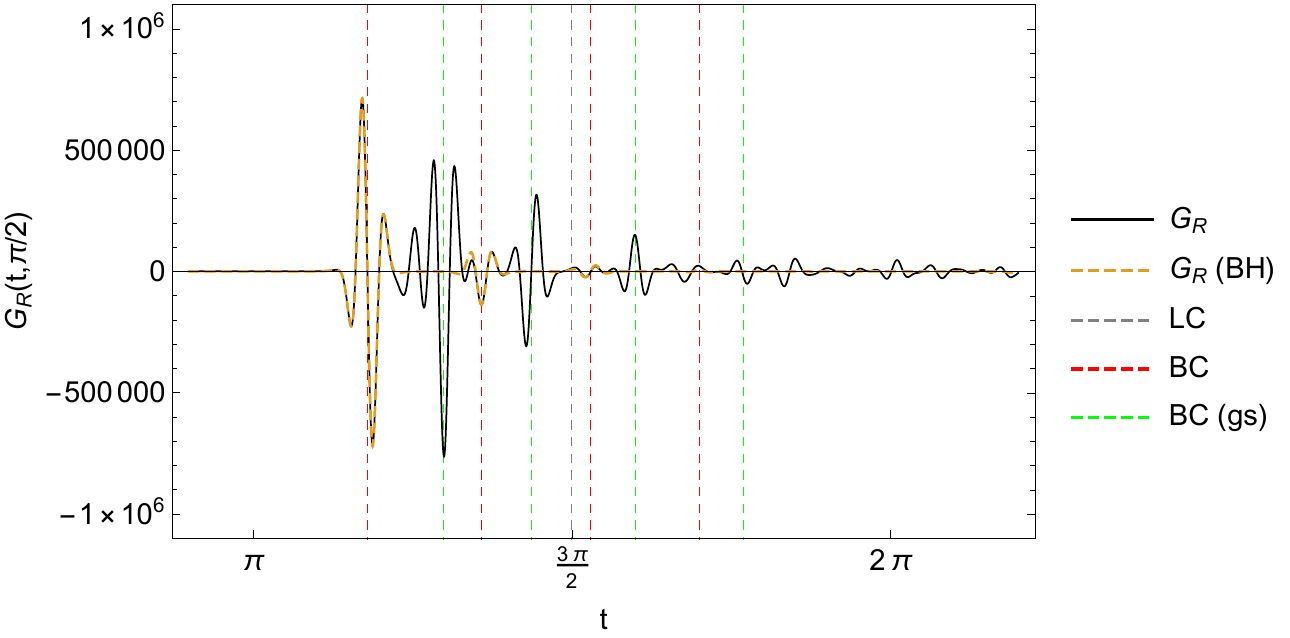}
\includegraphics[width=7.9cm]{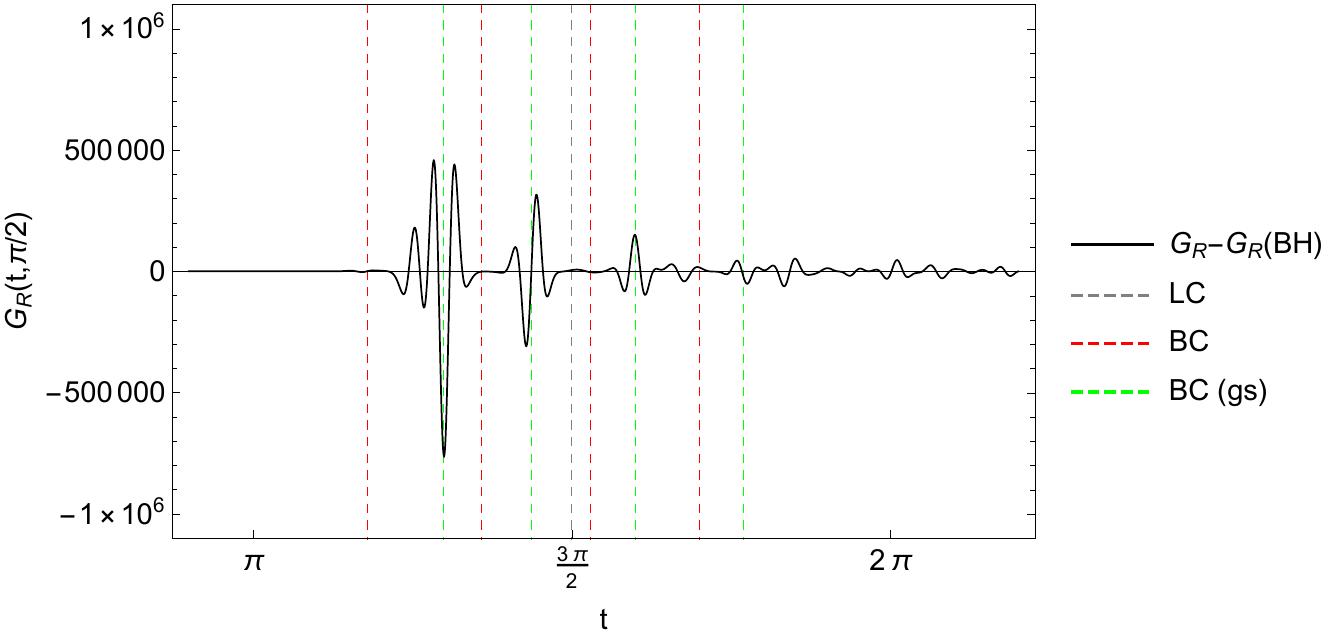}
\caption{The Green function $G_R(t,\pi/2)$ for the small gravastar in the region $\pi < t < 2\pi$. The right figure represents the Green function subtracted by that for black hole.}
\label{fig:GRt_GV_r0_007_01}
\end{figure}

\begin{figure}[ht]
\centering
\includegraphics[width=3.9cm]{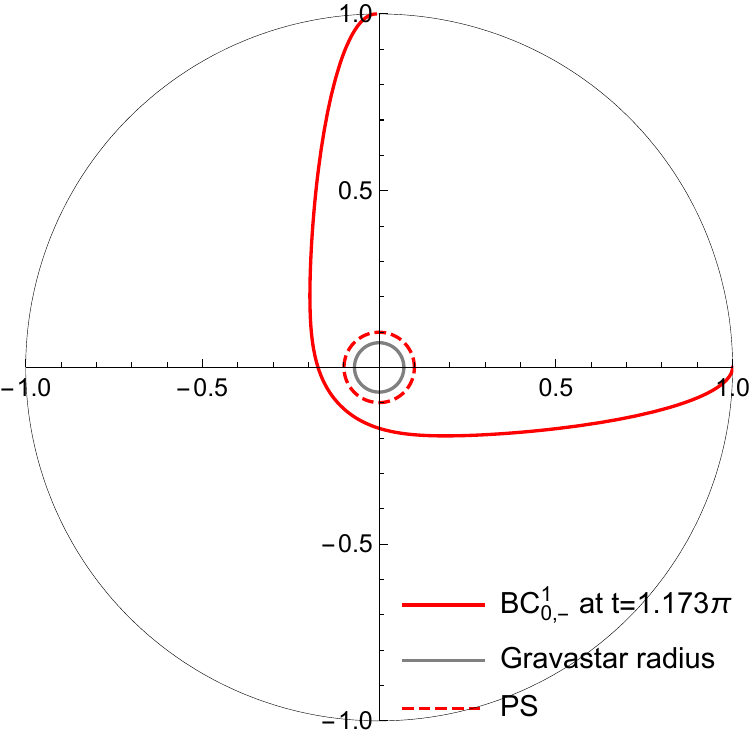}
\includegraphics[width=3.9cm]{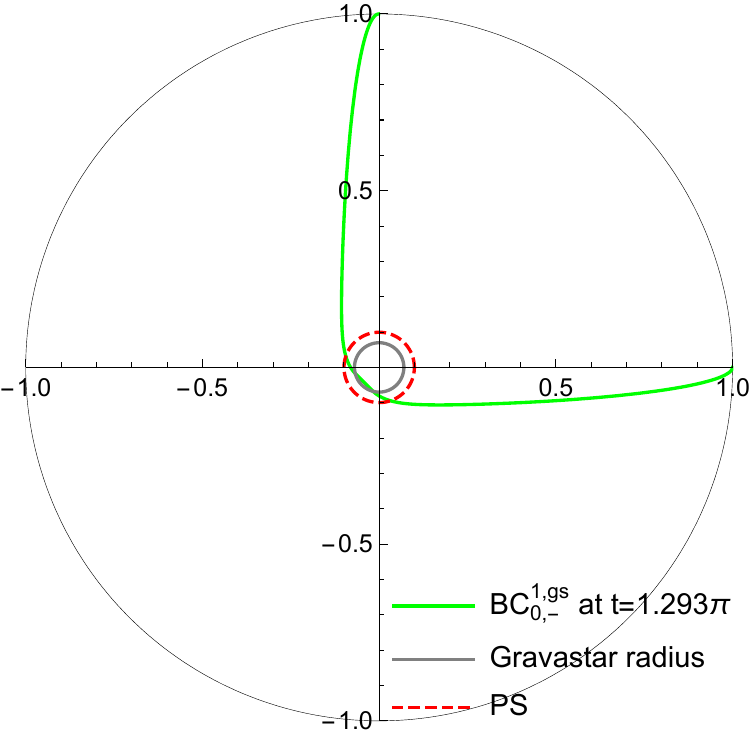}
\includegraphics[width=3.9cm]{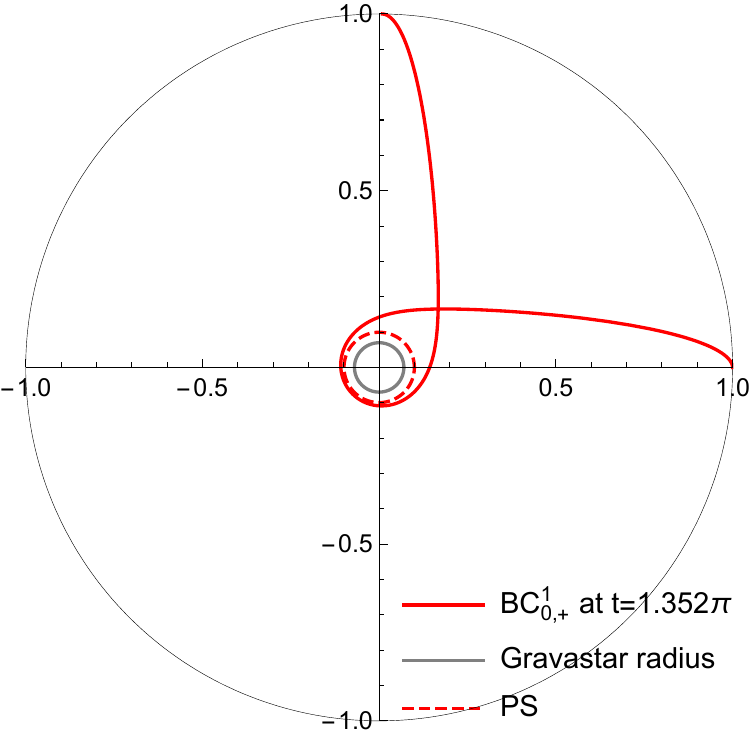}
\includegraphics[width=3.9cm]{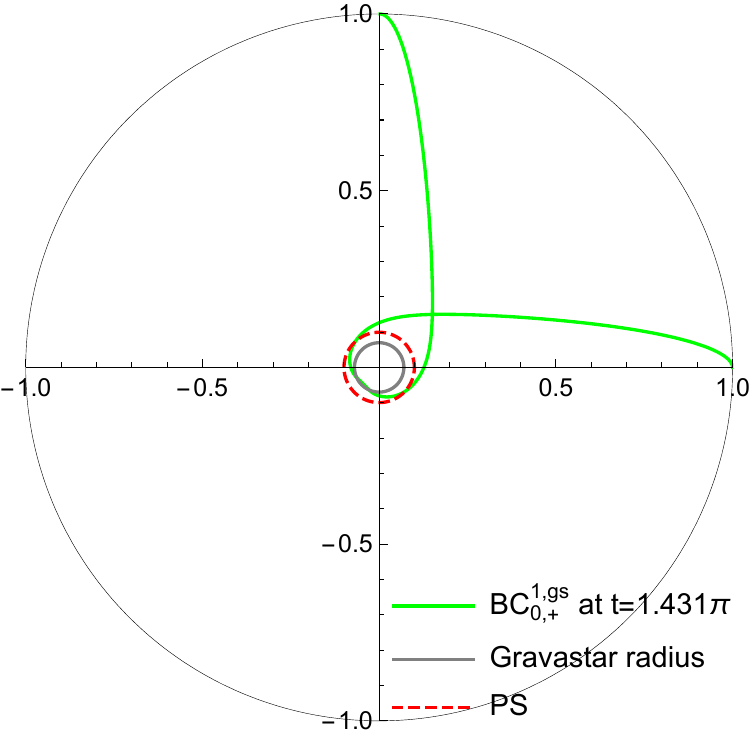}
\caption{Some null geodesic trajectories for the first series in the case of small gravastar. See also Fig.~\ref{fig:orbits-1stseries_r0_007} in Appendix~\ref{sec:trajectories} for the detail.}
\label{fig:orbits-1stseries_r0_007_few}
\end{figure}

\begin{figure}[ht]
\centering
\includegraphics[width=7.5cm]{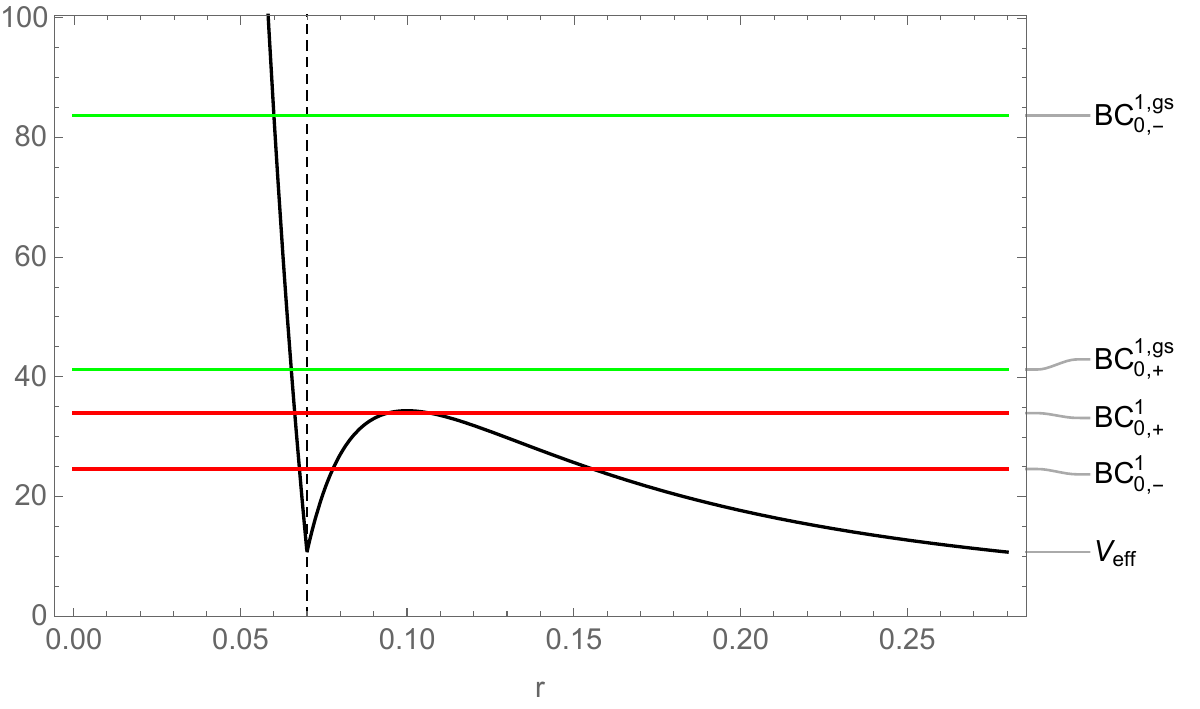}
\caption{Energy levels on the effective potential for some bulk cones of the first series for the small gravastar case. See also Fig.~\ref{fig:energylevels-1stseries_r0_007} in Appendix~\ref{sec:trajectories} for the detail.}
\label{fig:energylevels-1stseries_r0_007_few}
\end{figure}

\begin{figure}
\centering
\includegraphics[width=7.9cm]{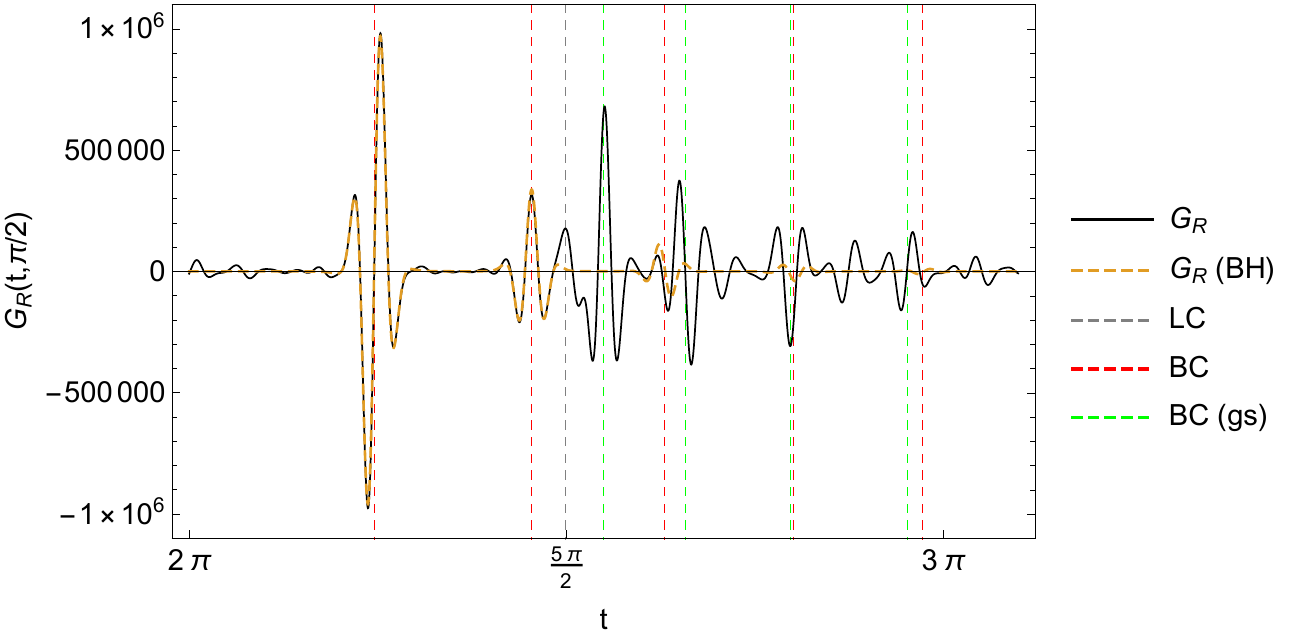}
\includegraphics[width=7.9cm]{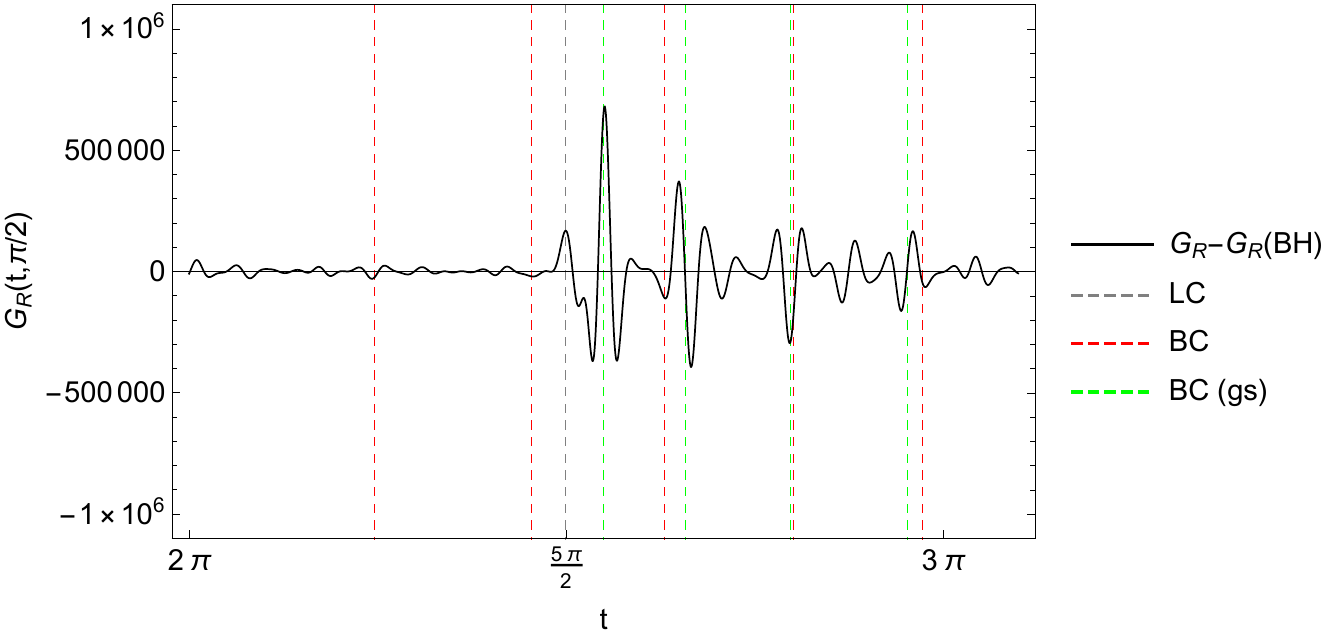}
\caption{The Green function $G_R(t,\pi/2)$ for the small gravastar in the region $2 \pi < t < 3\pi$. The right figure represents the Green function subtracted by that for black hole.}
\label{fig:GRt_GV_r0_007_02}
\end{figure}

For the numerical computation, we set the parameters as $R_\mathrm{AdS}=1$, $\mu=1/15$ and the shell radius $r_0=0.07$.
The photon sphere radius is $r_c=1/10\ (>r_0)$ and the horizon radius (if exists) is $r_h\simeq0.066$.
The result of numerical computation is summarized in Fig.~\ref{fig:GRt_GV_r0_007}.
As we will see in the following, there are two types of bumps. One type is common bumps to the black hole case and the other is characteristic bumps in the gravastar case.
The characteristic bumps are associated with the null geodesics passing through the gravastar interior.
The first series of bumps appears in $\pi<t<2\pi$ as in the left figure of Fig.~\ref{fig:GRt_GV_r0_007_01}.
The right figure in Fig.~\ref{fig:GRt_GV_r0_007_01} shows $G_R(t,\pi/2)$ with subtraction by $G_R(t,\pi/2)$ for black hole, i.e. the deviation, 
\begin{align}
(G_R(t,\pi/2) \text{ for gravastar})-(G_R(t,\pi/2) \text{ for black hole}).
\end{align}
The characteristic bumps are located at the arrival time expected from null geodesics passing through the gravastar interior.
The bulk-cones $\mathrm{BC}_{0,-}^{1,\mathrm{gs}}$, $\mathrm{BC}_{0,+}^{1,\mathrm{gs}}$, $\mathrm{BC}_{0,-}^{2,\mathrm{gs}}$, and $\mathrm{BC}_{0,+}^{2,\mathrm{gs}}$ appear in this time region.
We can see that the amplitudes of their bumps are of the same order as the black hole case.
Figs.~\ref{fig:orbits-1stseries_r0_007_few} and~\ref{fig:energylevels-1stseries_r0_007_few} briefly show the corresponding null geodesic trajectories and their energy levels on the effective potential $V_\mathrm{eff}(r)$, respectively, for the four earliest bulk-cones.
See Figs.~\ref{fig:orbits-1stseries_r0_007} and~\ref{fig:energylevels-1stseries_r0_007} in Appendix~\ref{sec:trajectories} for the details.
The second series appears in $2\pi<t<3\pi$ as in Fig.~\ref{fig:GRt_GV_r0_007_02}.
The green dashed lines represent the arrival times corresponding to the bulk-cones of $\mathrm{BC}_{1,+}^{1,\mathrm{gs}}$, $\mathrm{BC}_{1,-}^{2,\mathrm{gs}}$, $\mathrm{BC}_{1,+}^{2,\mathrm{gs}}$, and $\mathrm{BC}_{1,-}^{3,\mathrm{gs}}$.
The characteristic bumps still have  large amplitude, however, their appearance is delayed compared to the bumps associated with the photon sphere bulk-cones.
The corresponding trajectories and the energy levels are shown in Fig.~\ref{fig:orbits-2ndseries_r0_007} and the right figure of Fig.~\ref{fig:energylevels-1stseries_r0_007}.

We observe several interesting features from the numerical results, Figs.~\ref{fig:GRt_GV_r0_007}-\ref{fig:GRt_GV_r0_007_01}.
First, the four-fold structure of the bumps appears as in the case of the black hole~\cite{Dodelson:2020lal}.
In Figs.~\ref{fig:GRt_GV_r0_007_01}-\ref{fig:GRt_GV_r0_007_02}, the Green function after the subtraction, i.e., the sequence of the characteristic bumps of the gravastar, also exhibits the four-fold structure independently from that of the black hole case.
Second, the characteristic bumps in the gravastar case have the same order of amplitude as the black hole case.
This fact suggests that the strength of the bulk-cones in the gravastar case may be also related to the Lyapunov exponent of the photon sphere~\cite{Dodelson:2023nnr}.
In fact, some of the corresponding null geodesics have the energy levels close to the top of the effective potential at the photon sphere as shown in Fig.~\ref{fig:energylevels-1stseries_r0_007}.
Third, there are continuous weak signals other than the strong bumps.
These signals would be the waves that have been trapped in the potential well around the shell radius $r_0$ for a while and finally escape from it by tunneling (see Fig.~\ref{fig:energylevels-1stseries_r0_007}).
It is obvious that such waves cannot be prior to the first bulk-cone bumps, which are reflected by the outer potential barrier without being trapped in the well, and actually, weak signals do not appear before the first appearance of the strong bumps in Fig.~\ref{fig:GRt_GV_r0_007}.
Moreover, the continuous weak signals are not observed in the cases without potential well, see, e.g., Fig.~\ref{fig:GRt_BH} (and Fig.~\ref{fig:GRt_LGV_r0_017} below).
This phenomenon has also been observed in the context of the gravitational echo in asymptotically flat spacetimes~\cite{Cardoso:2016rao,Cardoso:2016oxy,Abedi:2016hgu,Oshita:2018fqu,Oshita:2020dox}.
The appearance of echoes is a common feature of perturbation around black hole alternatives, such as ultra-compact boson stars, (asymptotically flat) gravastars, and wormholes.

So far, we have investigated the gravastar with radius $r_0=0.07$ as an example.
It is worth noting that the smaller radius $r_0$ leads to shift of the bulk-cones $\mathrm{BC}_{n-1,\pm}^{j,\mathrm{gs}}$ to later time.
These bulk-cones have trajectories similar to those appearing in Figs.~\ref{fig:orbits-1stseries_r0_007} and~\ref{fig:orbits-2ndseries_r0_007}.
However, as the lapse function $f(r)$ around the radius $r_0$ becomes smaller for smaller $r_0$, the propagation time of the null geodesics across $r_0$ is elongated.
In particular, the delay should become arbitrary long in the limit $r_0\to r_h$.
Thus, we would be able to measure how small the gravastar is from the delay of appearance of the bulk-cones associated with null geodesics passing through the gravastar.

\subsection{Large gravastar}

\begin{figure}
\centering
\includegraphics[width=16cm]{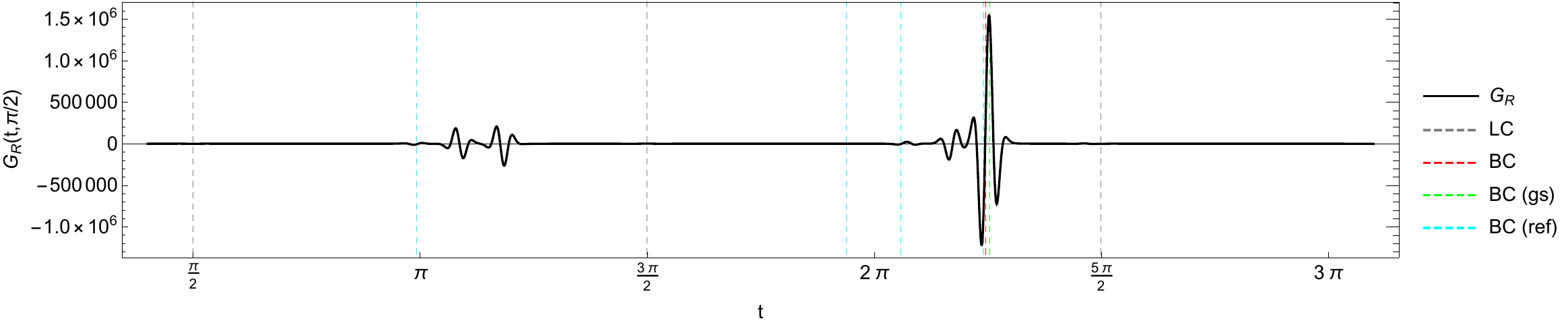}
\caption{The Green function $G_R(t,\pi/2)$ for the large gravastar. The vertical dashed lines indicate the arrival times of null geodesics.
Red, green, and cyan for the orbits reflected by the photon sphere potential, orbits transmitting the interior region, and fictitious orbits reflected by the shell, respectively.}
\label{fig:GRt_LGV_r0_017}
\end{figure}

The small gravastar has the photon sphere as in the black hole case. There are bulk-cone singularities associated with null geodesics largely affected by the photon sphere. We observed that these bulk-cone singularities exist in the both cases of small gravastar and black hole. The position of shell $r_0$ is a free parameter, and the small gravastar is realized by setting $r_c > r_0$ with the photon sphere radius $r_c$. As mentioned above, the large gravastar can be constructed by setting $r_c < r_0$. In this Subsection, we examine bulk-cone singularities for large gravastar.
We perform numerical computation by setting the gravastar shell radius as $r_0=0.17$, which is larger than the photon sphere radius $r_c=1/10$. The value of $r_c$ is extracted from the asymptotic parameters $R_\mathrm{AdS}=1$ and $\mu=1/15$. 
The result of numerical computation is summarized in Fig.~\ref{fig:GRt_LGV_r0_017}.
We can see that some of common bumps to the black hole case disappear in this case. 

\begin{figure}
\centering
\includegraphics[width=8cm]{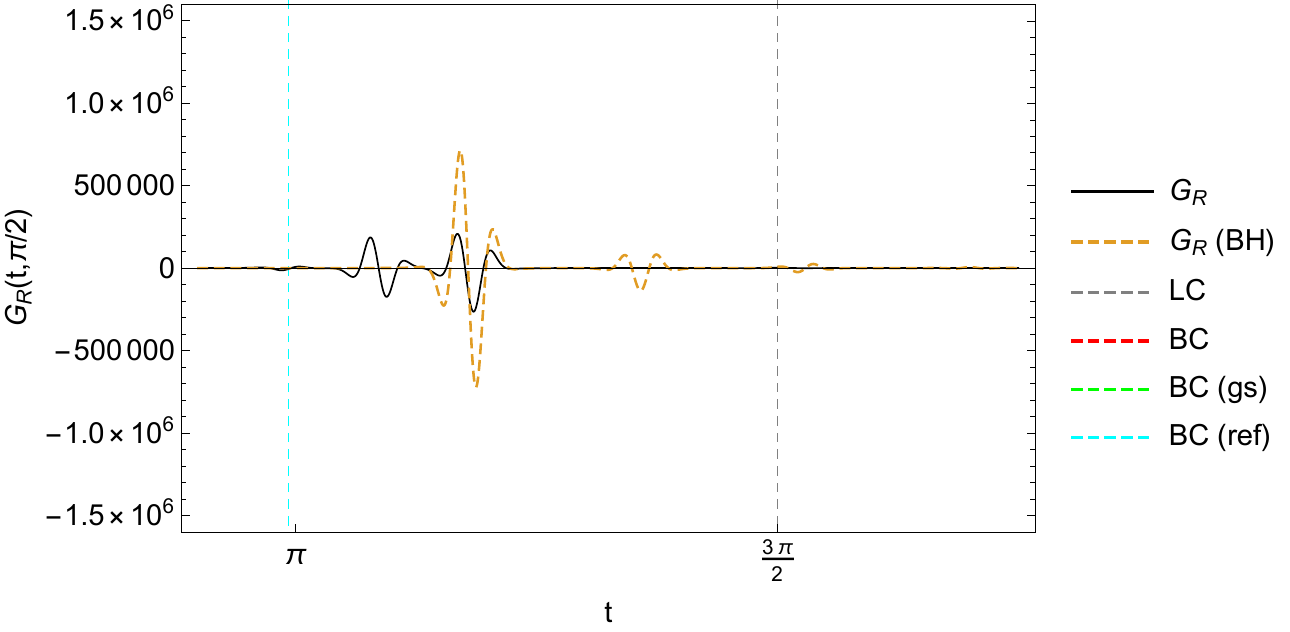}
\caption{The Green function $G_R(t,\pi/2)$ for the large gravastar in the region $ \pi < t < 3\pi/2$. }
\label{fig:GRt_LGV_r0_017_01}
\end{figure}

The first series of a few weak bumps appears in $\pi<t<3\pi/2$, see Fig.~\ref{fig:GRt_LGV_r0_017_01}.
In this time region, no bulk-cones are expected from the geodesic analysis.
One may expect that the weak bumps correspond to the waves reflected by the junction point $r=r_0$, at which the potential is non-smooth.%
\footnote{Imagine the Schr\"{o}dinger equation with a box potential. Even if the energy level is higher than the potential, a part of the wave is reflected at the step, accompanied with phase shift, or equivalently, time delay.}
The cyan dashed line in Fig.~\ref{fig:GRt_LGV_r0_017_01} represents the fictitious null geodesic that is reflected at $r=r_0$.
Here we may call this characteristic time the fictitious bulk-cone time.
The small bumps appear after this time.
The corresponding trajectory and the energy level for the potential are shown in Fig.~\ref{fig:orbits-1stseries_r0_017}.

\begin{figure}
\centering
\includegraphics[width=7.9cm]{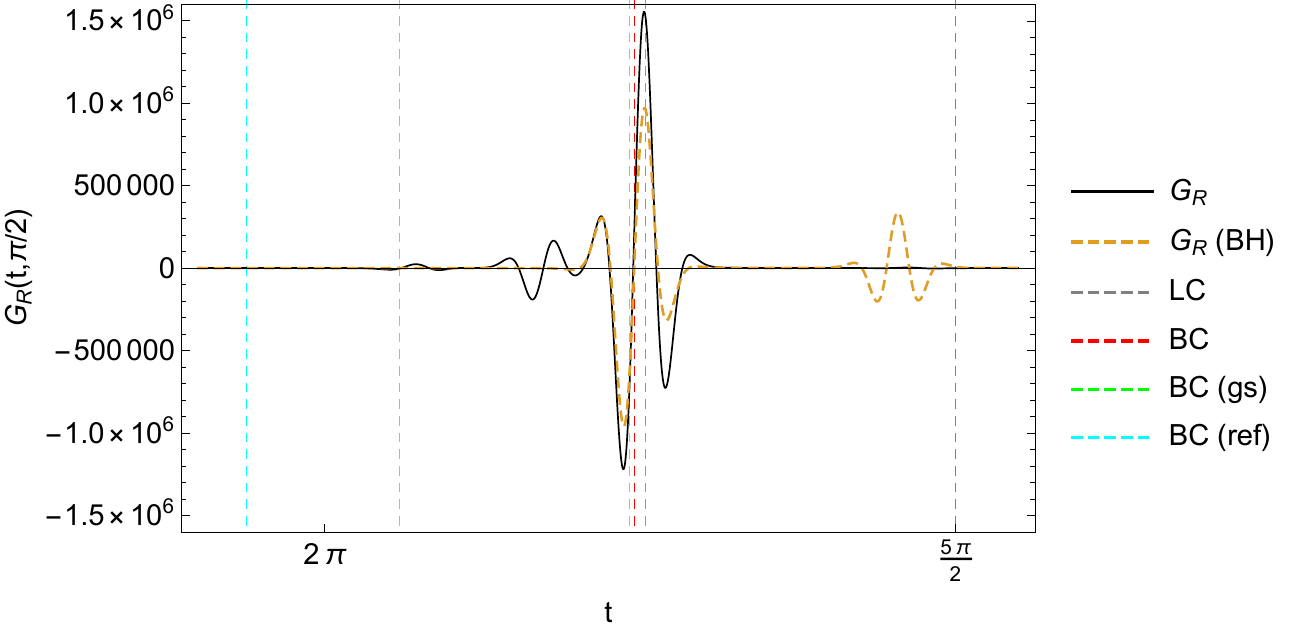}
\includegraphics[width=7.9cm]{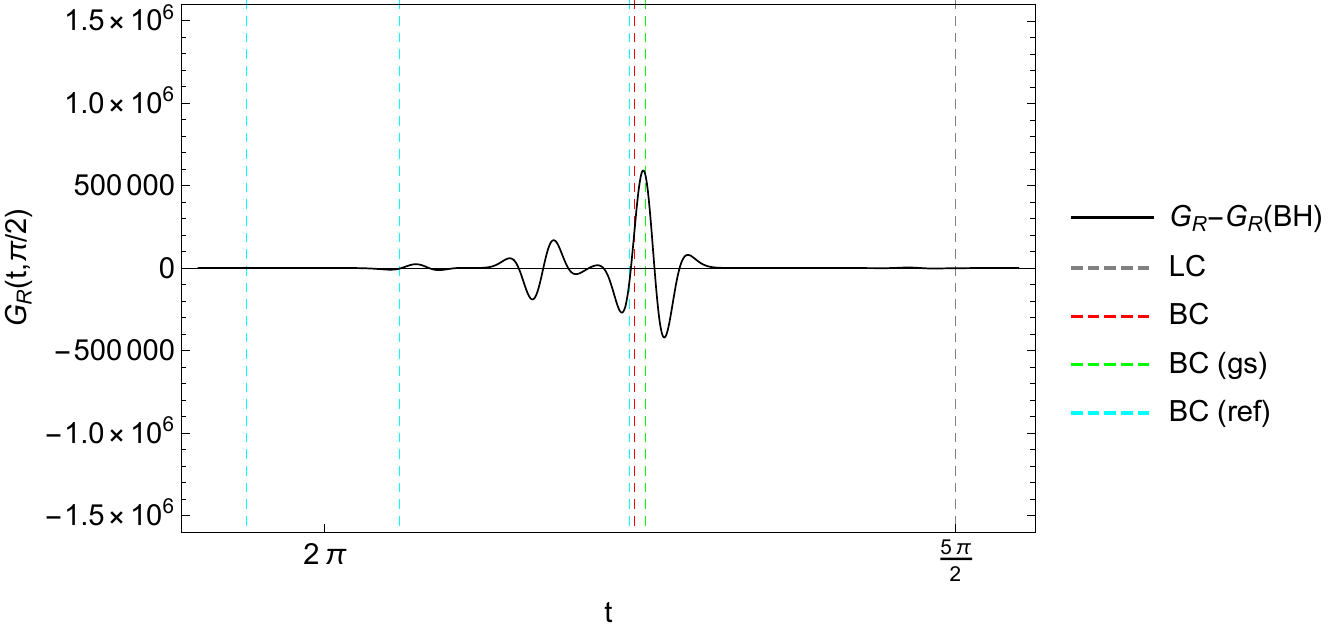}
\caption{The Green function $G_R(t,\pi/2)$ for the large gravastar in the region $2 \pi < t < 5\pi/2$. The right figure represents the Green function subtracted by that for black hole.}
\label{fig:GRt_LGV_r0_017_02}
\end{figure}

The second series of bumps appears in $2\pi<t<5\pi/2$.
The left figure of Fig.~\ref{fig:GRt_LGV_r0_017_02} is for the original $G_R(t,\pi/2)$.
There are bulk-cones of $\mathrm{BC}_{1,+}^{1}$ and $\mathrm{BC}_{1,+}^{1,\mathrm{gs}}$.
Although there is no photon sphere in the geometry, the null geodesic corresponding to $\mathrm{BC}_{1,+}^{1}$ exists because the turning point is even larger than the gravastar radius $r_0$.
The three fictitious bulk-cones are also depicted.
In the right figure of Fig.~\ref{fig:GRt_LGV_r0_017_02}, only the bump of the black hole case at $\mathrm{BC}_{1,+}^{1}$ is subtracted from $G_R(t,\pi/2)$.
The remaining bump is actually located at $\mathrm{BC}_{1,+}^{1,\mathrm{gs}}$.
The other small signals would be the waves reflected at the junction point as they appear after the fictitious bulk-cone times.
The corresponding trajectories and the energy levels on the effective potential are obtained as in Figs.~\ref{fig:orbits-2ndseries_r0_017} and~\ref{fig:energylevels-2ndseries_r0_017}, respectively.
We have numerically obtained the third series of bumps in $3\pi<t<7\pi/2$. We do not explain the detailed analysis here since nothing is significantly new compared with the case of second series.

\section{Conclusion and discussion}
\label{sec:conclusion}

In this note, we explicitly constructed the thin-shell model of AdS gravastar, which can be regarded as an alternative to black hole in asymptotically AdS spacetime. The exterior and interior regions are given by AdS-Schwarzschild black hole and dS spacetime. The two geometries are joined through a thin shell. We examined the null geodesics in such a geometry and divided them into two types. One type is the same as those of black hole, which goes around the photon sphere. Another type is specific to the gravastar case, and the null geodesics go into the gravastar interior region.
Assuming the existence of a holographic dual CFT on the boundary, we then numerically computed its retarded Green function from the wave functions of scalar field on the AdS gravastar. We observed the bumps corresponding to the null geodesics. In particular, we can identify the geometry as AdS gravastar by observing the bumps corresponding to the null trajectories going into the gravastar interior. We also studied the case of a large gravastar, where the shell is located outside the photon sphere. {We observed that several bulk-cones are absent in this case.} We also found the small bumps which may correspond to fictitious null geodesics reflected by the shell. In this way, we showed that the structure inside the photon sphere can be examined from the bulk-cone singularities in the retarded Green functions of dual CFT.

There are many open problems which we would like to consider. 
In this note, we have only considered the simplest model of a four-dimensional AdS gravastar, but it is possible to construct more general gravastar geometries in other dimensions.
One important problem is the stability of these geometries. It was claimed that the asymptotic flat gravastar is unstable due to nonlinear instabilities, see, e.g., \cite{Cardoso:2014sna}.
We hope the current setup would be helpful for the issue of nonlinear instabilities.
In particular, we would like to know the final state of the geometry due to the non-linear instability.
As the dual CFT is expected to be thermalized after some time, AdS/CFT correspondence suggests that the final geometry is black hole. It is also important to examine the cases of other black hole alternatives, such as, boson stars (see, e.g., \cite{Cardoso:2019rvt}). We are currently trying to reveal the generic features of bulk-cone singularities associated with black hole alternatives.

In this note, we numerically computed the retarded Green function from the wave functions of the bulk scalar field in the bulk geometry. It might be possible to express it in a semi-analytic way as in \cite{Dodelson:2023nnr} (see \cite{Dolan:2011fh} for the asymptotic flat case) by solving the wave equations in the WKB approximation. Several characteristic features of bulk singularities for a small gravastar have been observed in Section \ref{sec:smallgravastar}.
It is expected that we could explain these features by the WKB analysis. 
We hope to report our findings on the issue in near future.
It is also important to study long-lived modes associated with the potential minimal for the AdS gravastar, which is expected to be related to the non-linear instabilities of the geometry mentioned above.
See \cite{Cardoso:2014sna} for the asymptotic flat case.
The bulk-cone singularities exist only in strongly coupled CFTs and they are expected to be resolved in weakly coupled regimes. We would like to examine how the stringy or gravitational corrections resolve the bulk-cone singularities, see, e.g., \cite{Dodelson:2020lal,Dodelson:2023nnr}.
The bulk-cone singularities may be explained as instanton corrections in dual CFT, see \cite{Maldacena:2015iua} (and also \cite{Rey:2005cn,Rey:2006bz}).
It is also worthwhile to find out the string realization of the AdS gravastar.

\subsection*{Acknowledgements}

We are grateful to Koji Hashimoto, Takaaki Ishii, and Naritaka Oshita for useful discussions. The work of H.\,Y.\,C. is supported in part by Ministry of Science and Technology (MOST) through the grant 110-2112-M-002-006-. 
H.\, Y.\, C. would also like to thank YITP, Kyoto University for the hospitality where part of the work was carried out.
The work of Y.\,H. is supported by JSPS KAKENHI Grant Numbers JP21H05187 and JP23K25867. The work of Y.\,K. is supported by JSPS KAKENHI Grant Numbers JP21K20367 and JP23KK0048.

\appendix

\section{Technical details on AdS gravastar}
\label{app:AdSgravastar}

In this Appendix, we describe the detailed analysis on AdS gravastar and some concrete examples of null geodesics in the geometry.

\subsection{Thin-shell model}

\label{app:thinshell}

We start with the general construction of AdS gravastar metric.
We then move into the simplest case of thin-wall approximation, which we perform numerical analysis for the retarded Green function of dual CFT. 
Let $ds^2_\pm$ be the exterior ($+$) and interior ($-$) metrics as \eqref{eq:ansatz}.
We cut and join the two geometries at $r_\pm=r_0$.
The first junction condition requires that the induced metrics are equal, $h_+=h_-\equiv h$, and it gives the relation between the time coordinates \eqref{eq:cond_ind}.
The second junction condition is given by \eqref{eq:cond_energy}, which
has two nontrivial components,
\begin{align}
    \frac{2}{r_0}\left(g_+^{-1/2}-g_-^{-1/2}\right)&=-8\pi \sigma,\\
    -g_+^{-1/2}\left(\frac{1}{2}\frac{f'_+}{f_+}+\frac{1}{r_0}\right)
    +g_-^{-1/2}\left(\frac{1}{2}\frac{f'_-}{f_-}+\frac{1}{r_0}\right)
    &=-8\pi \bar p.
\end{align}
In our case, the metric coefficients are given as \eqref{eq:fpfm} with positive constants $\Lambda_\pm>0$.

Following the analysis~\cite{Cardoso:2014sna}, let us construct the simplest model of AdS gravastar with the assumption $\sigma=0$, or so-called thin-shell model.
The second junction condition reduces to
\begin{align}
    \frac{2}{r_0}\left(\sqrt{f_+}-\sqrt{f_-}\right)&=0,\\
    -\sqrt{f_+}\left(\frac{1}{2}\frac{f_+'}{f_+}+\frac{1}{r_0}\right)
    +\sqrt{f_-}\left(\frac{1}{2}\frac{f_-'}{f_-}+\frac{1}{r_0}\right)
    &=-8\pi \bar p,
\end{align}
where the quantities are evaluated at $r=r_0$.
The first equation results in
\begin{align}
    0=f_+(r_0)-f_-(r_0)=\frac{\Lambda_++\Lambda_-}{3}r_0^2-\frac{\nu}{r_0},
\end{align}
which leads to
\begin{align}
    \label{eq:r0}
    r_0=\left(\frac{3 \mu}{\Lambda_++\Lambda_-}\right)^{1/3}.
\end{align}
Substituting the expression, it is useful to rewrite
 the metric factors as
\begin{align}
    f_+(r_0)=f_-(r_0)=1-\frac{\Lambda_-}{3}\left(\frac{3 \mu}{\Lambda_++\Lambda_-}\right)^{2/3}.
\end{align}
The second equation, with the above result, determines the tension as
\begin{align}
    8\pi \bar p=\frac{1}{2}f_+^{-1/2}\left(f_+'-f_-'\right)
    =\frac{3}{2}\mu^{1/3}\left(\frac{\Lambda_++\Lambda_-}{3}\right)^{2/3}\left[1-\frac{\Lambda_-}{3}\left(\frac{3 \mu}{\Lambda_++\Lambda_-}\right)^{2/3}\right]^{-1/2}.
\end{align}
Note that for the pressure to be real valued, the parameters must satisfy the inequality,
\begin{align}
    \frac{\Lambda_-}{3}\left(\frac{3 \mu}{\Lambda_++\Lambda_-}\right)^{2/3}<1,
\end{align}
or equivalently,
\begin{align}
    \label{eq:real-pressure}
    \mu<\sqrt{3} \left( \frac{\Lambda_++\Lambda_-}{\Lambda_-^{3/2}} \right)=:\mu_\mathrm{max}.
\end{align}
In summary, the thin-shell model of AdS gravastar with $\sigma=0$ is specified by the three parameters, $\{\Lambda_+,\Lambda_-,\mu\}$, with the joint boundary $r=r_0$ determined by Eq.~\eqref{eq:r0}.
Since $r_0$ depends monotonically on each parameter, we can alternatively choose the three independent parameters as, for example, $\{\Lambda_+,\Lambda_-,r_0\}$ with $\mu$ given from Eq.~\eqref{eq:r0}.

To obtain the desired model, we further need to restrict the parameters.
Let us work with the parameter set $\{\Lambda_+,\Lambda_-,r_0\}$.
The horizon-less condition must be imposed,
\begin{align}
    r_0>r_h&:=\{r|f_+(r)=0\}
\end{align}
where $r_h$ is the horizon radius of the original AdS-Schwarzschild geometry as in \eqref{eq:horizon}.
Since $f_+(r)>0$ if and only if $r>r_h$, the condition is equivalent to
\begin{align}
\label{eq:horizon-less-cond2}
    f_+(r_0)>0.
\end{align}
We may also require that the gravastar is so small that it has a photon sphere outside the surface (see \eqref{eq:ps-cond})
\begin{align}
\label{eq:ps-cond2}
    r_0<r_c:=\{r|(f_+(r)r^{-2})'=0\}=\frac{3}{2} \mu.
\end{align}
Furthermore, the resulting parameter $\mu=\mu(\Lambda_+,\Lambda_-,r_0)$ should satisfy the real pressure condition~\eqref{eq:real-pressure}.
Since $r_0$ monotonically increases as $\mu$ does, the upper bound $\mu_\mathrm{max}$ for $\mu$ gives that for $r_0$,
\begin{align}
\label{eq:c-horizon-less-cond}
    r_0<r_\mathrm{max}:=
    \left(\frac{3 \mu_\mathrm{max}}{\Lambda_++\Lambda_-}\right)^{1/3}=\sqrt{\frac{3}{\Lambda_-}}.
\end{align}
This is equivalent to the condition that the joint boundary for the interior geometry is inside the cosmological horizon.
The first two of the above inequalities, Eqs.~\eqref{eq:horizon-less-cond2}--\eqref{eq:c-horizon-less-cond}, depend on $\mu$, which is now a redundant parameter with $r_0$.
Using Eq.~\eqref{eq:r0}, Eq.~\eqref{eq:horizon-less-cond2}  is written as
\begin{align}
    f_+(r_0)=1-\frac{\Lambda_-}{3}r_0^2>0.
\end{align}
This is equivalent to Eq.~\eqref{eq:c-horizon-less-cond} and Eq.~\eqref{eq:ps-cond2} reduces to 
\begin{align}
    r_0>\sqrt{\frac{2}{\Lambda_++\Lambda_-}}.
\end{align}
Finally, the condition for the AdS gravastar to be a horizon-less compact object with a photon sphere is
\begin{align}
\label{eq:r0-condition}
    \sqrt{\frac{2}{\Lambda_++\Lambda_-}}
    <r_0
    <\sqrt{\frac{3}{\Lambda_-}}.
\end{align}
Clearly, the left-hand side of the inequality is always smaller than the right-hand side and there always exists  such $r_0$.

\subsection{Shell dynamics}
\label{sec:shell-dynamics}

Here we examine the perturbative stability of AdS gravastar. The following calculation is based on Ref.~\cite{Kokubu:2014vwa}.
We consider a dynamical spherical shell obtained from cut and paste of two manifolds $\mathcal{M}_\pm$ with the metrics,
\begin{align}
    ds_\pm^2=-f_\pm(r)dt^2_\pm+f_\pm(r)^{-1}dr^2+r^2d\Omega^2.
\end{align}
We assume that these spacetimes are $\Lambda$-vacuum solutions with possibly different cosmological constants.
The manifolds are cut at a time-dependent radius $r=a(\tau)$,
where $\tau$ is the proper time of the shell.
The parts $r_+\ge a$ and $r_-\le a$ are joined at the radius.
The shell is identified with the hypersurface $F:=r-a(\tau)=0$ and its induced metric is given as 
\begin{align}
    d\sigma^2=-d\tau^2+a(\tau)^2d\Omega^2.
\end{align}
The 4-velocity of the shell on each side is given by
\begin{align}
    u_\pm^\mu=\left(\frac{dt_\pm}{d\tau},\frac{da}{d\tau},0,0\right)
    =\left(f_\pm^{-1}\sqrt{f_\pm+\left(\frac{da}{d\tau}\right)^2},\frac{da}{d\tau},0,0\right).
\end{align}
Outward unit normals to the boundary are
\begin{align}
    (n_\pm)_\mu
    =\left(-\frac{da}{d\tau},f_\pm^{-1}\sqrt{f_\pm+\left(\frac{da}{d\tau}\right)^2},0,0\right).
\end{align}
The second junction condition can be written as
\begin{align}
    S^i_j=-\frac{1}{8\pi}\left(\kappa^i_j-\delta^i_j\kappa^l_l\right), \quad
    \kappa^i_j:=(\chi^+)^i_j-(\chi^-)^i_j.
\end{align}
Here $(\chi^\pm)^i_j$ is the extrinsic curvature, and the indices $i,j$ are for the coordinates $\xi^i$ on the shell.
The extrinsic curvature is explicitly given by
\begin{align}
    (\chi^\pm)^\tau_\tau=\frac{B_\pm}{A_\pm},\quad
    (\chi^\pm)^\theta_\theta=(\chi^\pm)^\phi_\phi=\frac{A_\pm}{a},
\end{align}
where 
\begin{align}
    A_\pm=\sqrt{f_\pm+\left(\frac{da}{d\tau}\right)^2},\quad
    B_\pm=\frac{d^2a}{d\tau^2}+\frac{1}{2}\frac{df_\pm}{dr}.
\end{align}
From the symmetry of the spacetime, the stress energy tensor of the shell must take the perfect fluid form, $S^i_j=\mathrm{diag}(-\sigma,\bar p,\bar p)$.
The independent components of the junction condition, $(\tau,\tau)$ and $(\theta,\theta)$, are then obtained as
\begin{align}
    \label{eq:jc-tautau}
    &\sigma=-\frac{1}{4\pi a}(A_+-A_-),\\
    \label{eq:jc-thetatheta}
    &\bar p=\frac{1}{8\pi}\left[\frac{B_+}{A_+}-\frac{B_-}{A_-}+\frac{1}{a}(A_+-A_-)\right].
\end{align}
In addition, we have the conservation law of the shell,
\begin{align}
    S^{ij}_{|j}+[[T^\alpha_\beta (e^i)_\alpha n^\beta ]]=0,
\end{align}
where $T_\pm$ is the energy momentum tensor of the spacetimes $ds^2_\pm$, $(e^i)_\alpha=\partial \xi^i/\partial x^\alpha$,
and $[[X]]=X_+-X_-$ denotes the jump of a quantity $X$ across the shell.
Since $ds_\pm^2$ are the metrics of $\Lambda$-vacuum solutions, the second term in the conservation law vanishes.
The remaining nontrivial equation of the conservation law is
\begin{align}
    \label{eq:consv-shell}
    \frac{d}{d\tau}(a^2\sigma) +\bar p\frac{d}{d\tau}(a^2)=0.
\end{align}
One of Eqs.~\eqref{eq:jc-tautau},~\eqref{eq:jc-thetatheta}, and~\eqref{eq:consv-shell} is reducible from the other two.
The conservation law can be integrated once the EOS of the shell, $\bar p=\bar p(\sigma)$, is specified.

Let us suppose $\sigma=0$.
This assumption leads to
\begin{align}
    a(\tau)=\mathrm{const.}
\end{align}
That is, the shell cannot be dynamical in this simplest model.
We suppose $\sigma\neq 0$ in the following.
Squaring Eq.~\eqref{eq:jc-tautau} twice, we obtain
\begin{align}
    &\left(\frac{da}{d\tau}\right)^2 +U=0,\\
    &U:=-\frac{1}{4}(4\pi a \sigma)^2-\frac{1}{4(4\pi a\sigma)^2}(f_+-f_-)^2+\frac{1}{2}(f_++f_-).
\end{align}
Specifying the EOS $\bar p=\bar p(\sigma)$ and solving Eq.~\eqref{eq:consv-shell}, we obtain $\sigma=\sigma(a)$.
Inserting it into the above equation, we obtain $U=U(a)$ and the problem reduces to the one-dimensional potential one, $\left(\frac{da}{d\tau}\right)^2+U(a)=0$.
Thus, a static shell is obtained by solving $U(a)=\frac{d}{da}U(a)=0$
and its radial stability is determined by the sign of $\frac{d^2}{da^2}U(a)$.

\subsection{Null geodesic trajectories}
\label{sec:trajectories}

Here we collect the figures describing the null geodesics appearing in the main text.
The radial part of null geodesic equations can be put into the form \eqref{eq:geodpotential}. We also indicate the energy levels with respective to the
effective potential $V_\text{eff}(r)$ for the null geodesics.

\begin{figure}
\centering
\includegraphics[width=3.9cm]{graphics_num/GV_r0_007/GV_r0_007_psorbit_01.pdf}
\includegraphics[width=3.9cm]{graphics_num/GV_r0_007/GV_r0_007_innerorbit_01.pdf}
\includegraphics[width=3.9cm]{graphics_num/GV_r0_007/GV_r0_007_psorbit_02.pdf}
\includegraphics[width=3.9cm]{graphics_num/GV_r0_007/GV_r0_007_innerorbit_02.pdf}\\
\includegraphics[width=3.9cm]{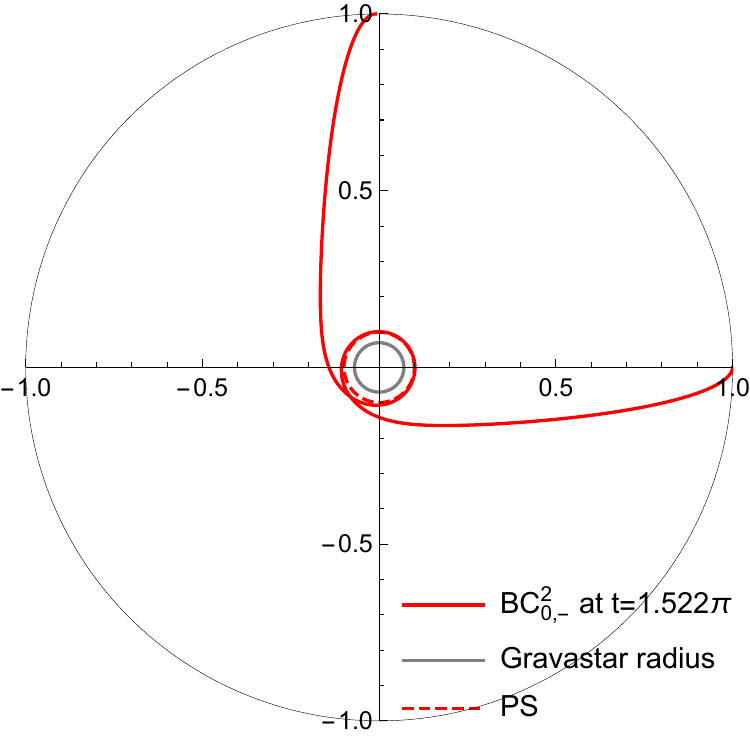}
\includegraphics[width=3.9cm]{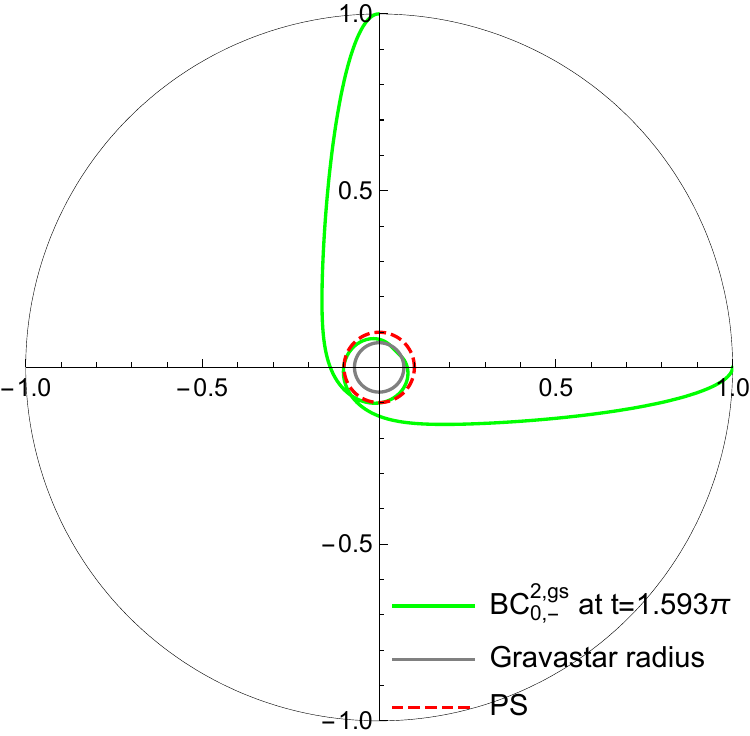}
\includegraphics[width=3.9cm]{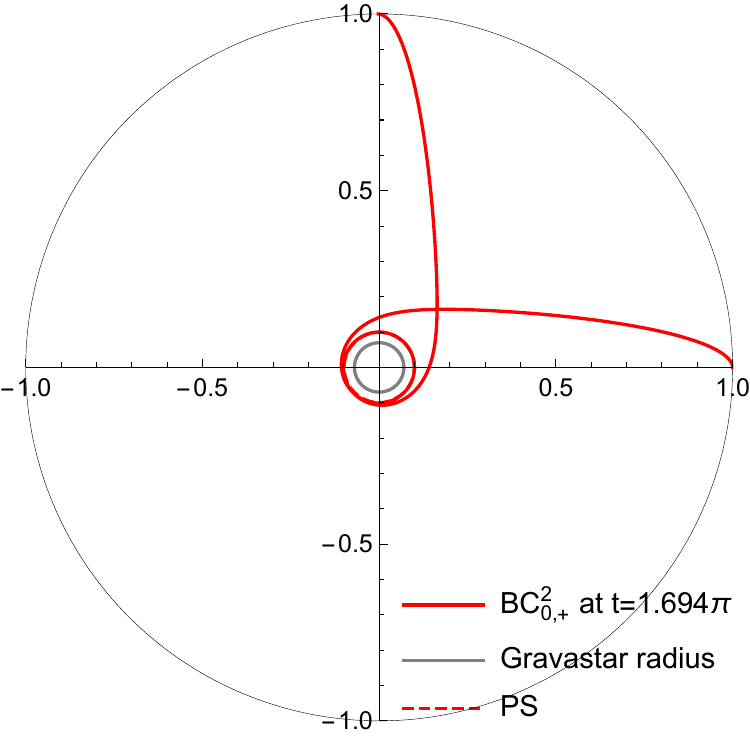}
\includegraphics[width=3.9cm]{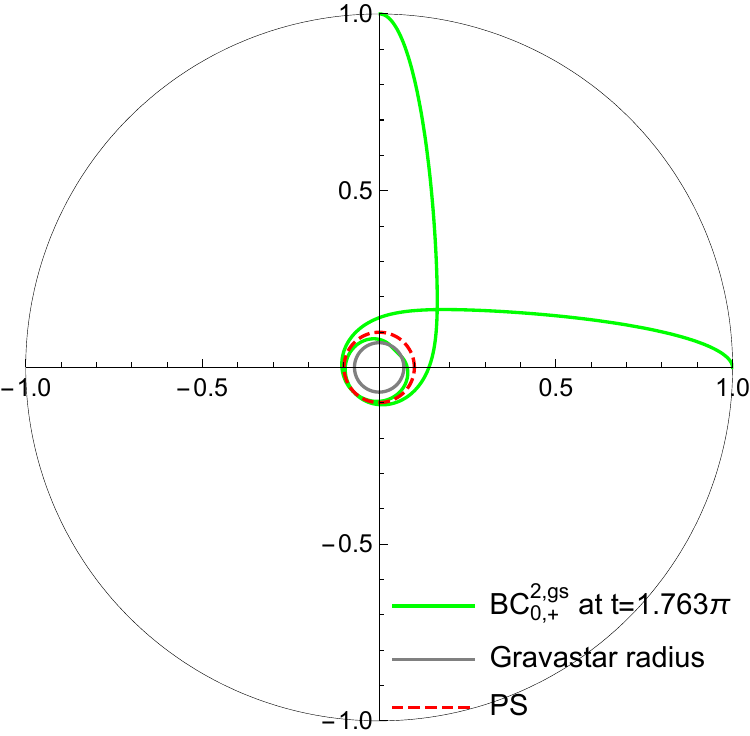}
\caption{Null geodesic trajectories for the first series in the case of small gravastar.}
\label{fig:orbits-1stseries_r0_007}
\end{figure}

\begin{figure}[ht]
\centering
\includegraphics[width=3.9cm]{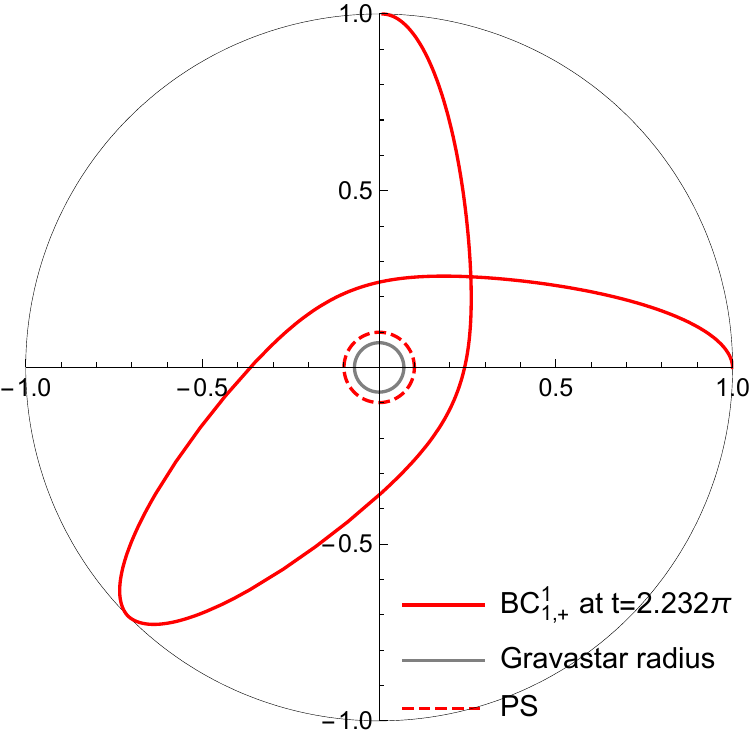}
\includegraphics[width=3.9cm]{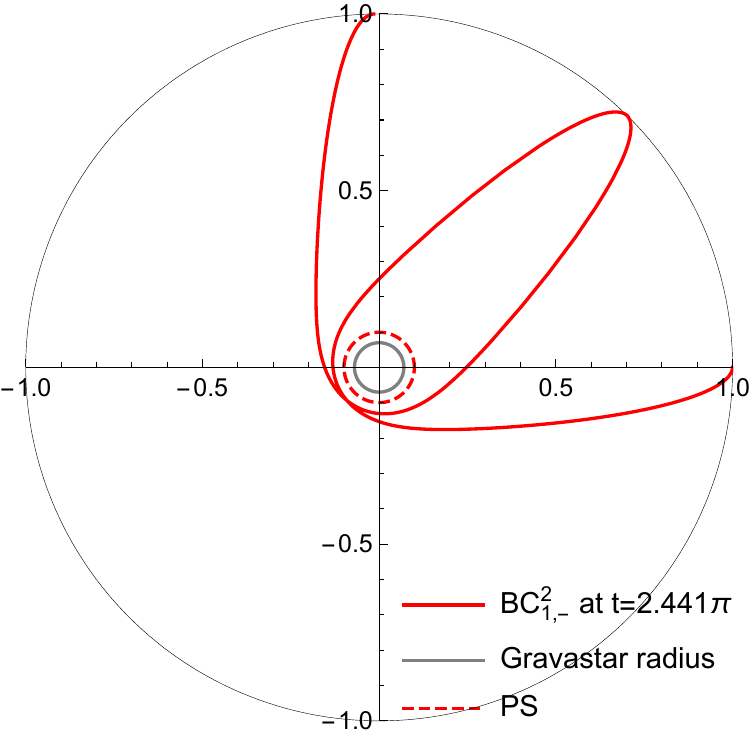}
\includegraphics[width=3.9cm]{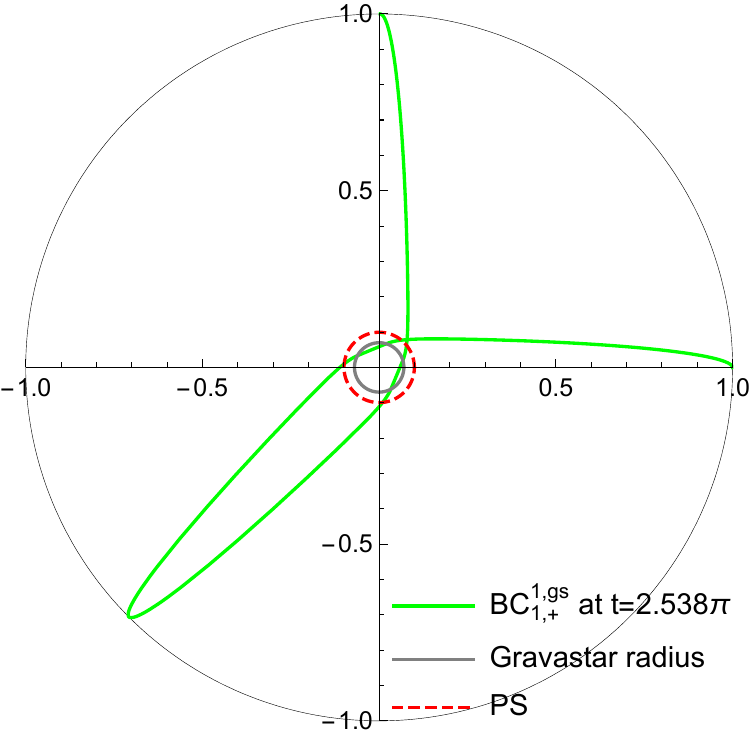}
\includegraphics[width=3.9cm]{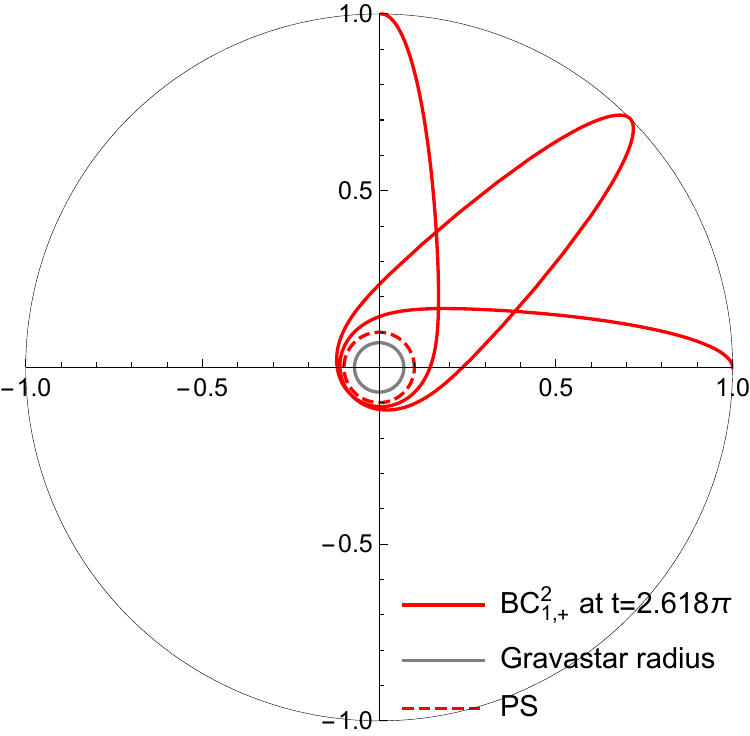}
\includegraphics[width=3.9cm]{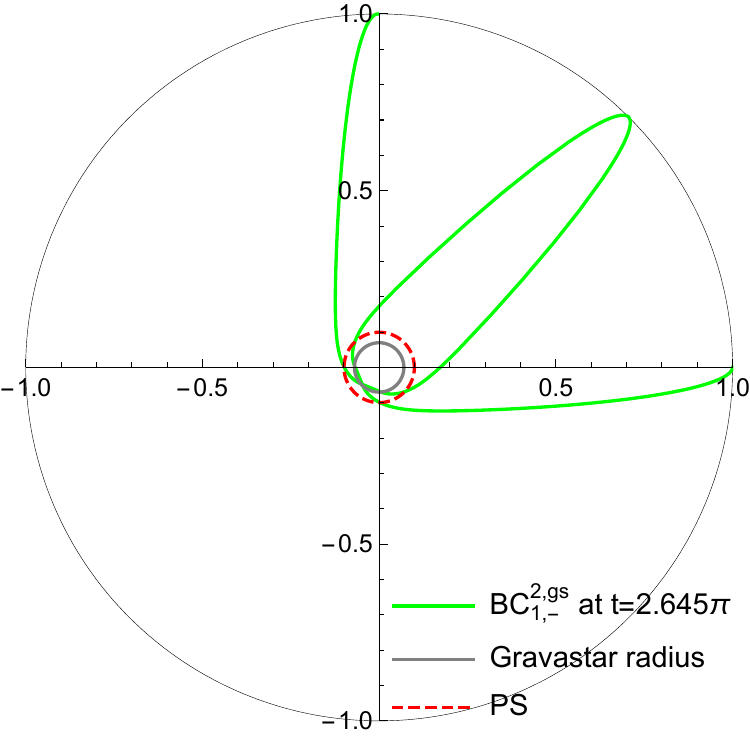}
\includegraphics[width=3.9cm]{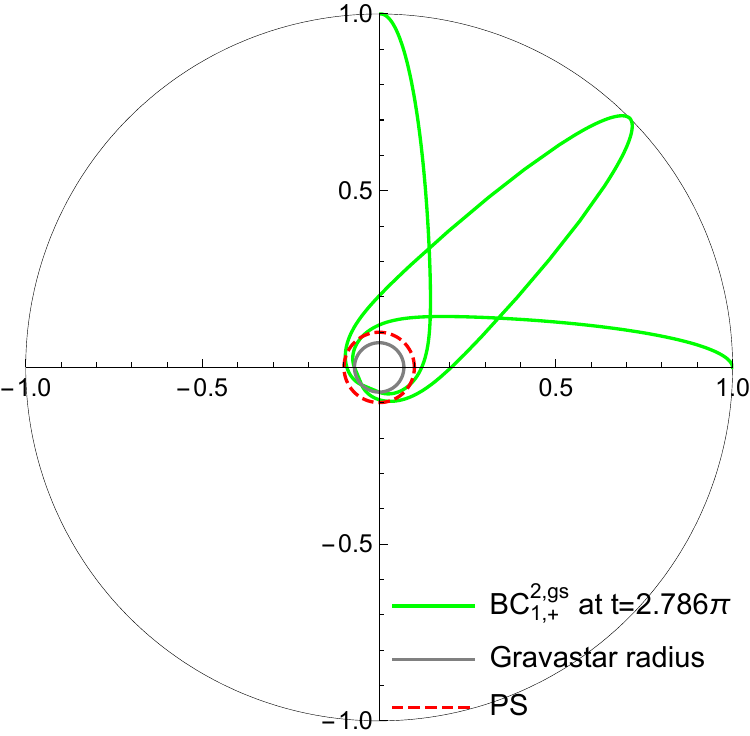}
\includegraphics[width=3.9cm]{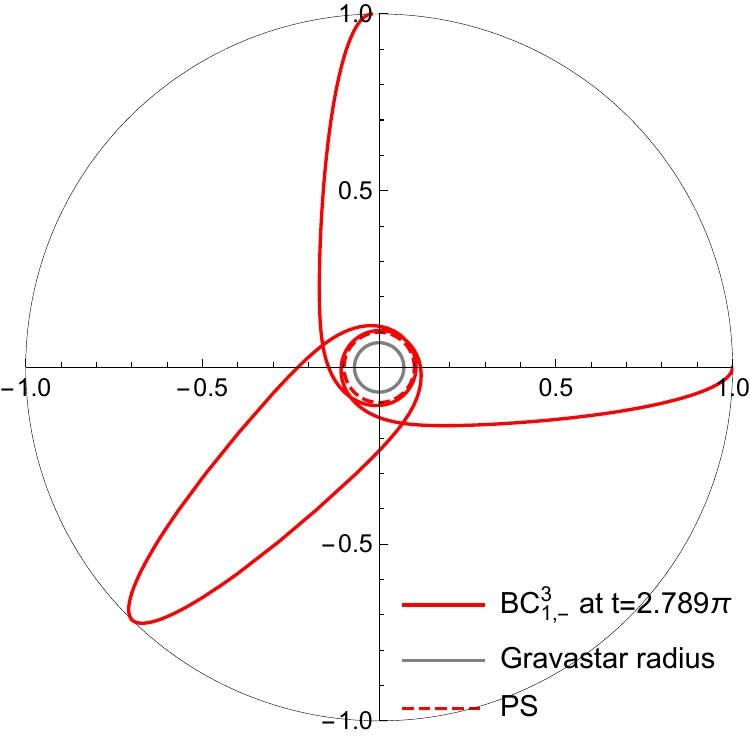}
\includegraphics[width=3.9cm]{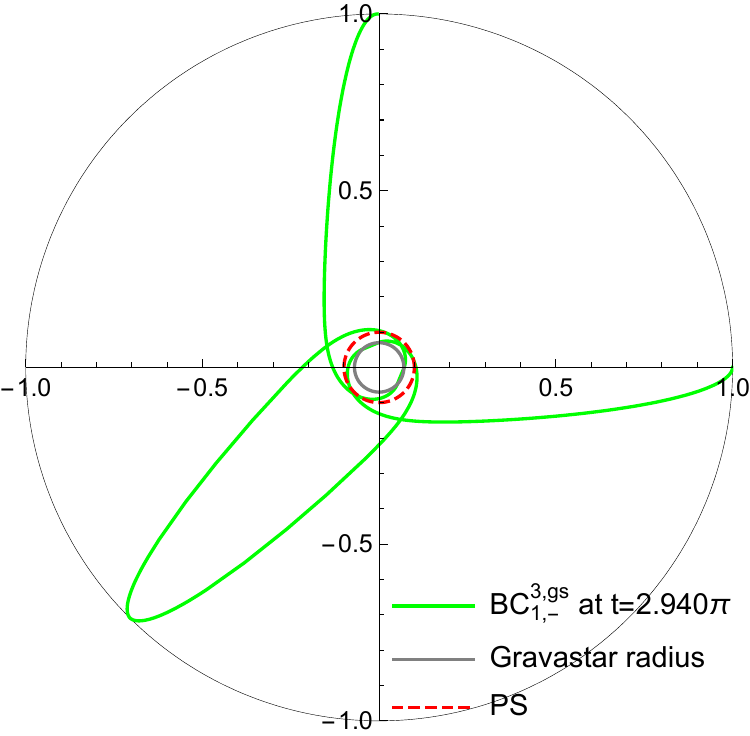}
\caption{Null geodesic trajectories for the second series  in the case of small gravastar.}
\label{fig:orbits-2ndseries_r0_007}
\end{figure}

\begin{figure}
\centering
\includegraphics[width=7.5cm]{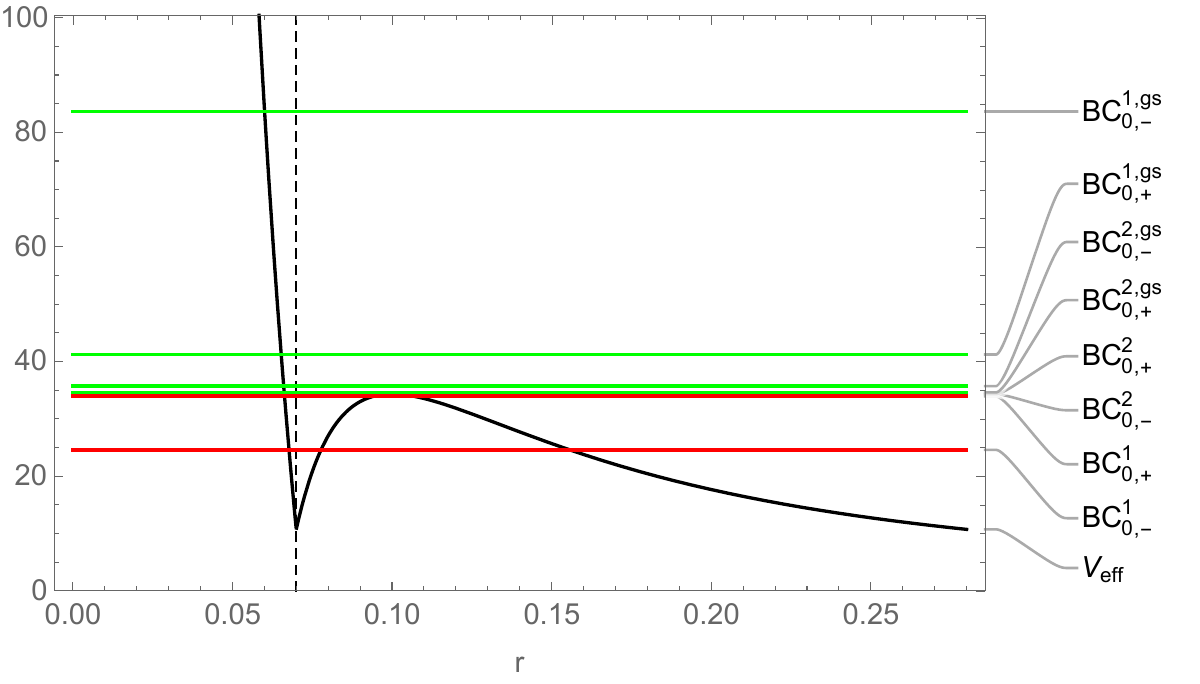}
\includegraphics[width=7.5cm]{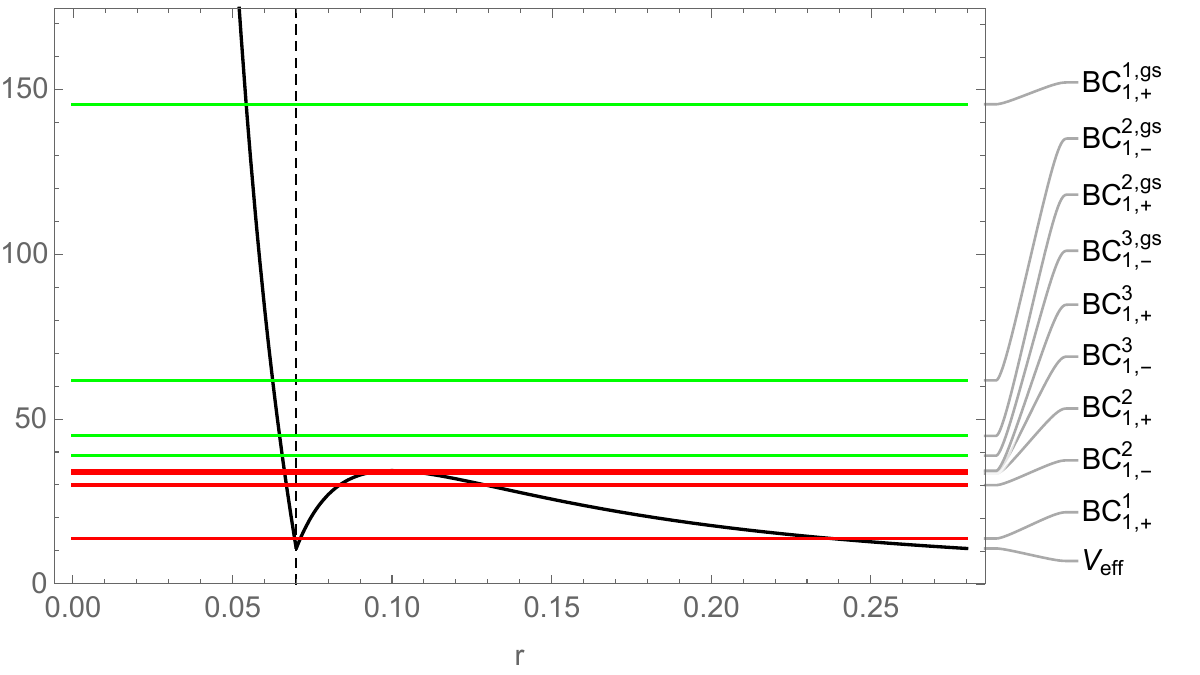}
\caption{Energy levels on the effective potential for the first series (Left) and for the second series (Right) in the case of small gravastar.}
\label{fig:energylevels-1stseries_r0_007}
\end{figure}

We first consider null geodesics in the case of a small gravastar with a photon sphere.
The first series of bumps appears in $\pi<t<2\pi$ as in Fig.~\ref{fig:GRt_GV_r0_007_01}.
The null geodesic trajectories for the bulk-cones are obtained as in Fig.~\ref{fig:orbits-1stseries_r0_007}.
The energy levels with respective to the effective potential are described in the left figure of Fig.~\ref{fig:energylevels-1stseries_r0_007}.
The second series appears in $2\pi<t<3\pi$ as in Fig.~\ref{fig:GRt_GV_r0_007_02}.
The corresponding trajectories and the energy levels with respective to the effective potential are shown in Fig.~\ref{fig:orbits-2ndseries_r0_007} and the right figure of Fig.~\ref{fig:energylevels-1stseries_r0_007}, respectively.

\begin{figure}
\begin{minipage}[t]{0.495\textwidth}
\centering
\includegraphics[width=3.9cm]{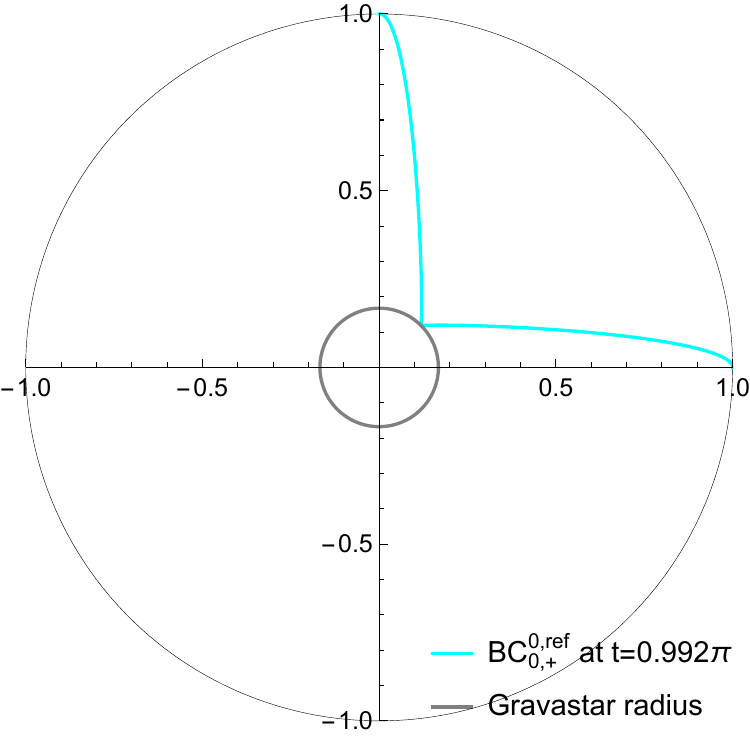}
\end{minipage}
\begin{minipage}[t]{0.495\textwidth}
\centering
\includegraphics[width=7.9cm]{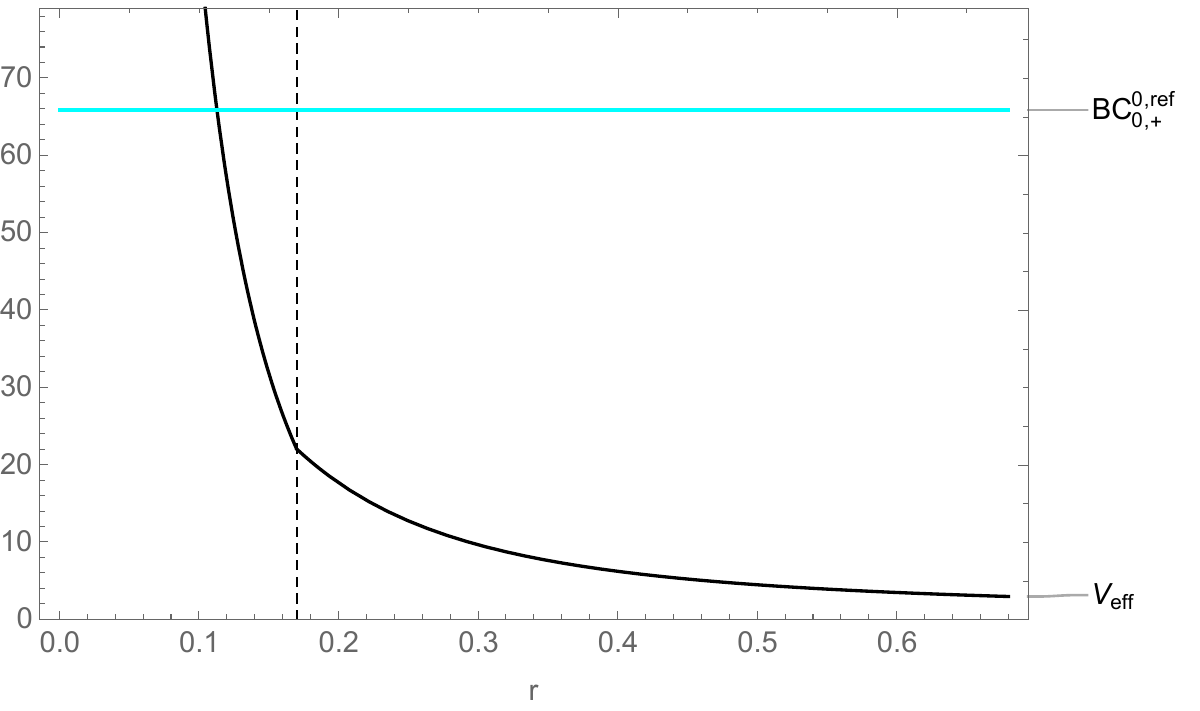}
\end{minipage}
\caption{Null geodesic trajectories (Left) and energy levels on the effective potential (Right) for the first series  in the case of large gravastar.}
\label{fig:orbits-1stseries_r0_017}
\end{figure}

\begin{figure}[H]
\centering
\includegraphics[width=3.9cm]{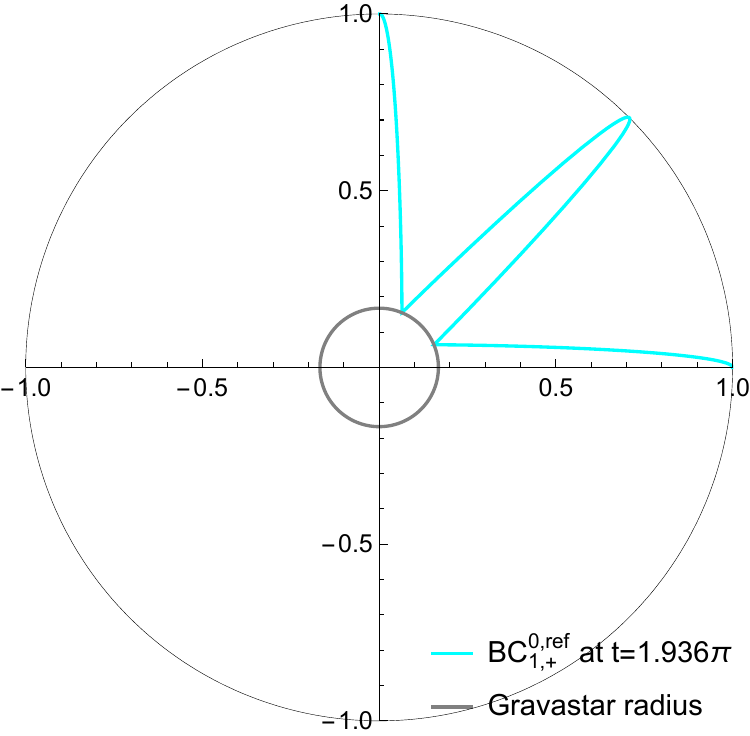}
\includegraphics[width=3.9cm]{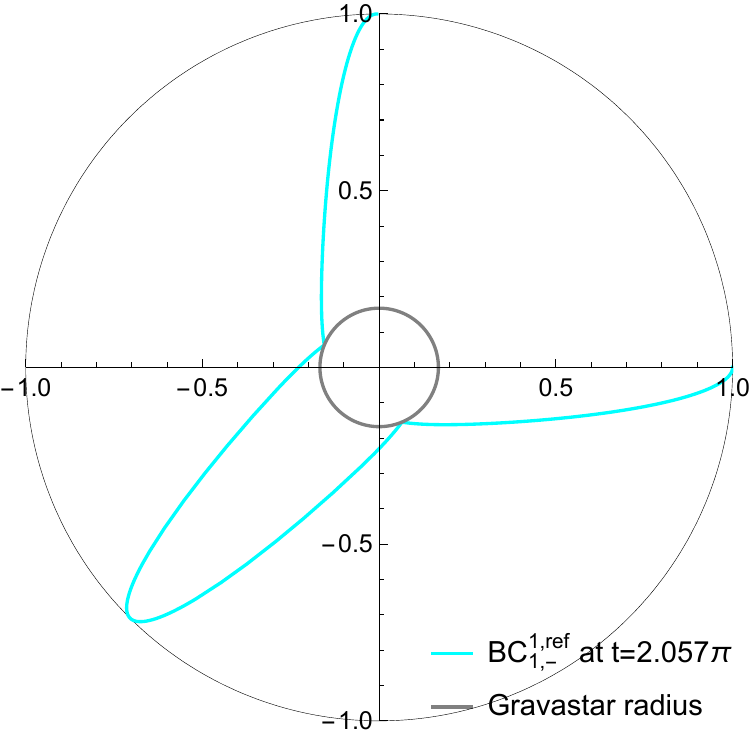}
\includegraphics[width=3.9cm]{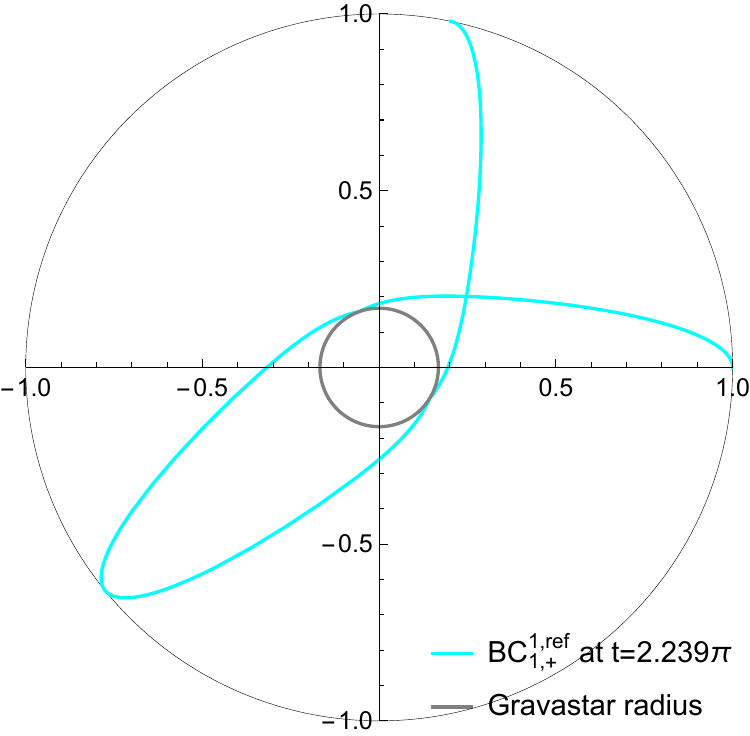}\\
\includegraphics[width=3.9cm]{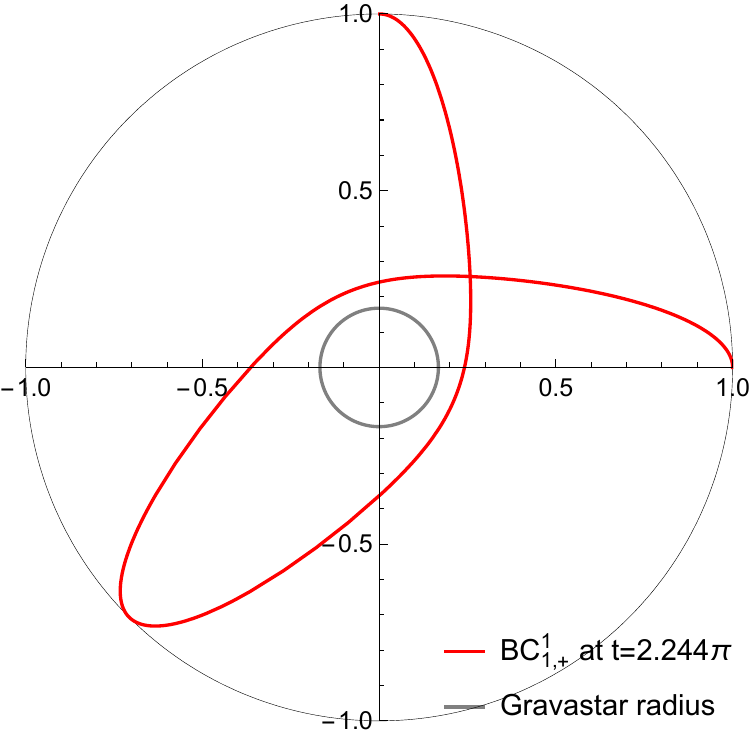}
\includegraphics[width=3.9cm]{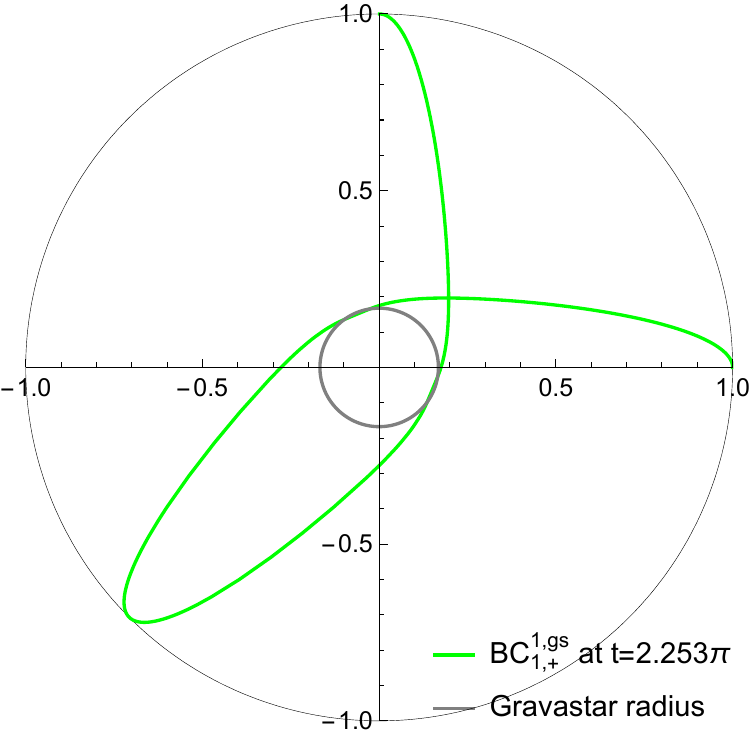}
\caption{Null geodesic trajectories for the second series in the case of large gravastar.}
\label{fig:orbits-2ndseries_r0_017}
\end{figure}

\begin{figure}[H]
\centering
\includegraphics[width=8cm]{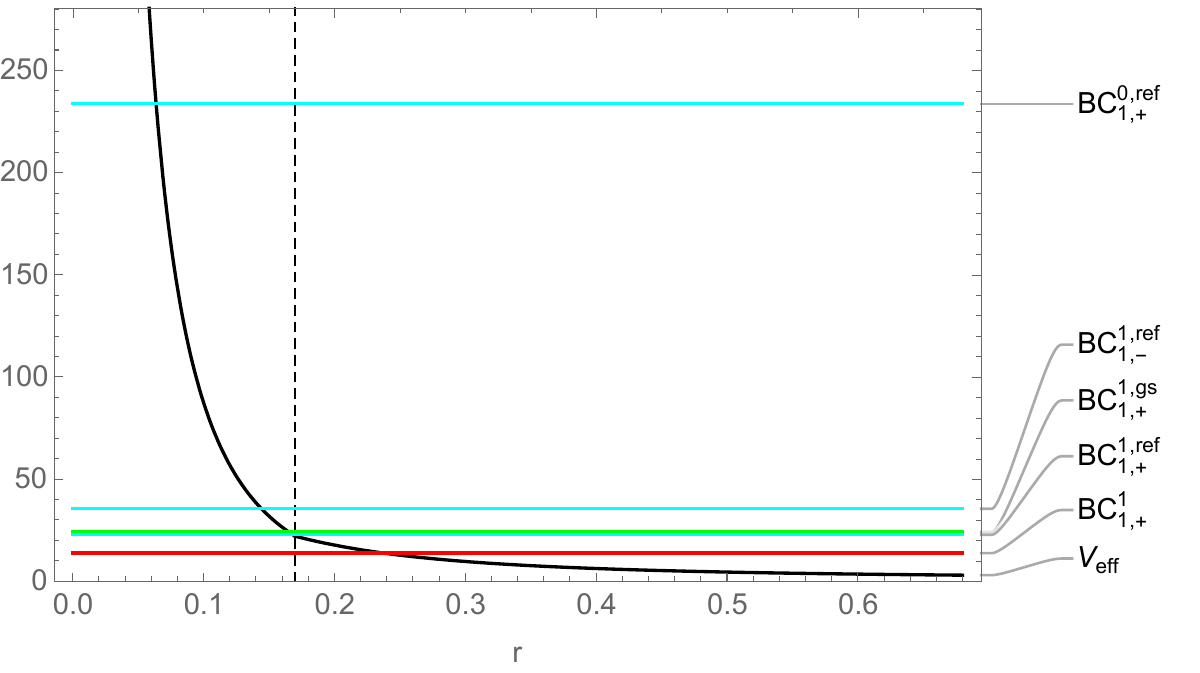}
\includegraphics[width=7.6cm]{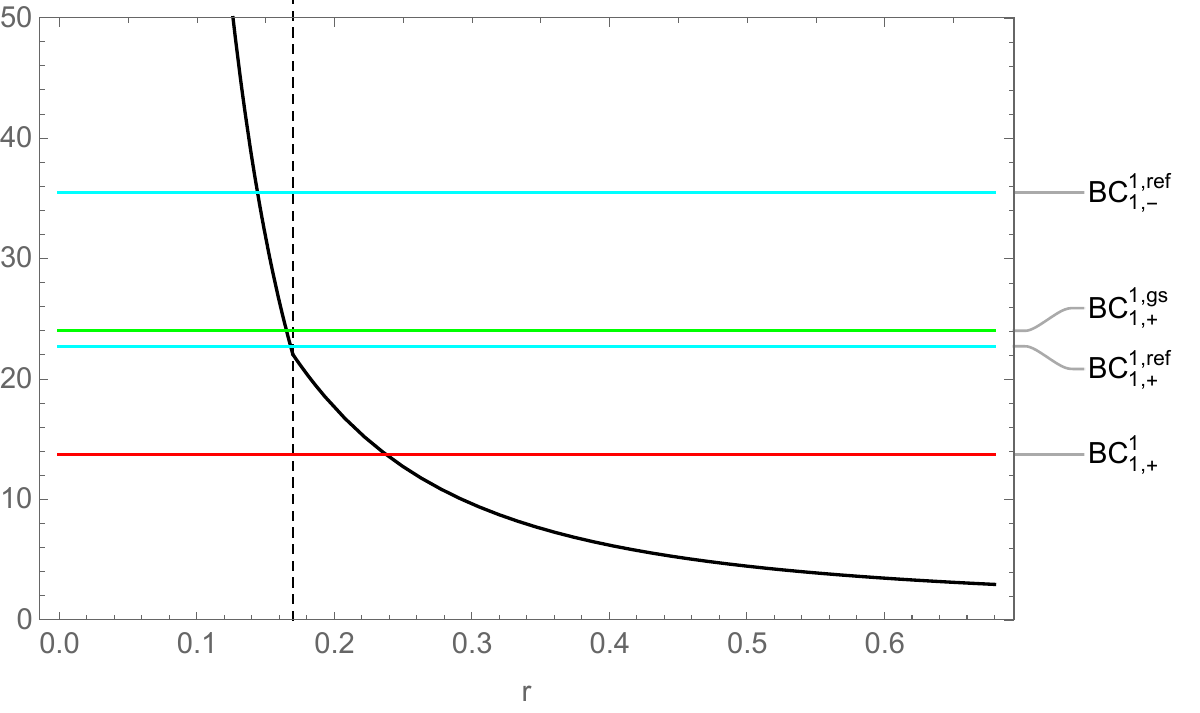}
\caption{Energy levels on the effective potential for the second series in the case of large gravastar.}
\label{fig:energylevels-2ndseries_r0_017}
\end{figure}

We then move to a large gravastar that does not have a photon sphere.
The first series of a few weak bumps appears in $\pi<t<3\pi/2$  as in Fig.~\ref{fig:GRt_LGV_r0_017_01}.
The corresponding trajectory and the energy level with respective to the effective potential are shown in Fig.~\ref{fig:orbits-1stseries_r0_017}.
The second series of bumps appears in $2\pi<t<5\pi/2$.
The corresponding trajectories and energy levels with respect to the effective potential are obtained as in Figs.~\ref{fig:orbits-2ndseries_r0_017} and~\ref{fig:energylevels-2ndseries_r0_017}, respectively.

\section{Analysis of wave equations}
\label{app:wave}

In this Appendix, we summarize some technical details on the analysis of wave equations.

\subsection{Eikonal potentials in the AdS gravastar}
\label{app:Eikonal}

In the main text, we studied the bulk-cone singularities of retarded Green functions from the wave functions of bulk scalar field and discussed the relation to bulk null geodesics.
We shall see the relation more directly by taking the eikonal limit for the wave equation of bulk scalar field.

In the case of AdS gravastar, the wave equation is given by \eqref{eq:zwaveeqp} with the potential \eqref{eq:gspotential}.
The eikonal limit is obtained by taking a large integer spin $\ell$.
In order to realize this, we take the limit $p\to\infty$ for $\ell+1/2=p$ and $\omega=pu$.
The ratio of the potential to $p^2$ is
\begin{align}
    \frac{V(z)}{p^2}=f(r)\left[\frac{1-\frac{1}{4p^2}}{r^2}+\frac{\nu^2-\frac{9}{4}}{p^2}+\frac{\frac{d}{dr}f(r)/r}{p^2}\right].
\end{align}
For large $p$, the potential in the range $r\ll p$ reduces as follows.
The first term in the square brackets reduces to $1/r^2$.
The second term is ignored compared to the first term because it is of order $1/p^2$.
The third term is also dropped as follows.
Outside the gravastar, we have $r\gtrsim \mu$ and the numerator becomes $\frac{d}{dr}f(r)/r=2+\frac{\mu}{r^3}\lesssim 2+r^{-2}$.
So, the third term is of order $1/p^2$ and dropped.
Inside the gravastar, the numerator is $\frac{d}{dr}f(r)/r=2R_\mathrm{dS}^{-2}$ and the third term is of order $1/(R_\mathrm{dS}p)^2\ll 1/R_\mathrm{dS}^2$. 
On the other hand, the first term is $1/r^2> 1/R_\mathrm{dS}^2$ since here $r<R_\mathrm{dS}$.
Thus, the third term is much smaller than the first term and is dropped.
Then we have the eikonal wave equation,
\begin{align}
    \left(\partial_z^2+p^2\kappa^2(z)\right)\psi(z)=0,
    \;\;\kappa(z)=\sqrt{u^2-V_\mathrm{eik}^{\ell+\alpha=p}(z)},
\end{align}
where
\begin{align}
    \frac{V(z)}{p^2}\simeq f(r)r^{-2}=:V_\mathrm{eik}^{\ell+\alpha=p} (z).
\end{align}
This potential coincides that of the null geodesic~\eqref{eq:geodpotential}.

\begin{figure}[ht]
\centering
\includegraphics[width=7.9cm]{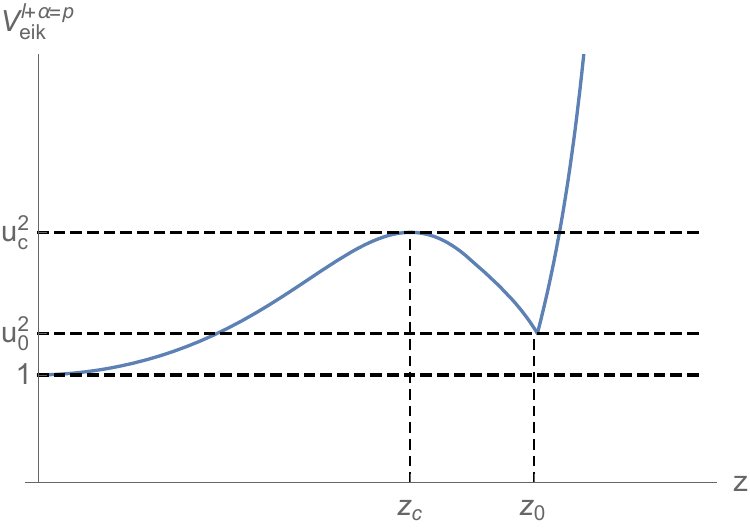}
\includegraphics[width=7.9cm]{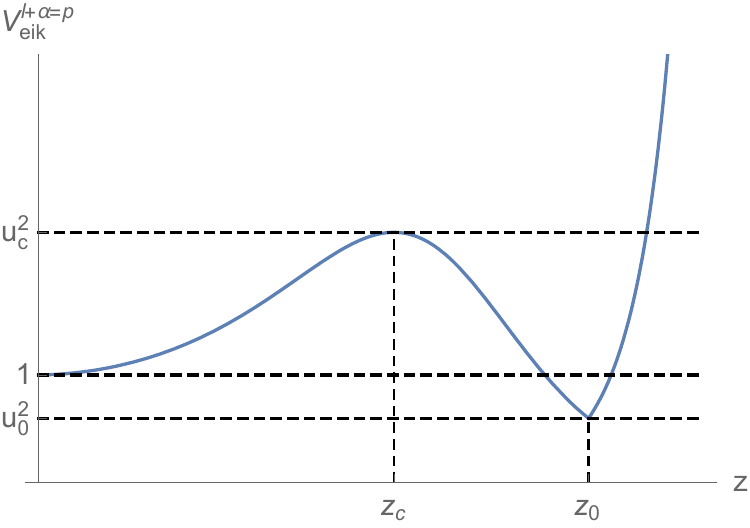}
\caption{Two types of the eikonal potential $V_\mathrm{eik}^{l+\alpha=p}$ for the AdS gravastar.}
\label{fig:eikonalpotential-gravastar-realspin}
\end{figure}

For a typical case, the eikonal potential is depicted in the left figure of Fig.~\ref{fig:eikonalpotential-gravastar-realspin}.
For $u^2>u_c^2$, 
the mode hits the potential once 
near the center.
For $1<u^2<u_c^2$, the mode hits the potential three times.
Note that the gravastar center corresponds to a positive finite value of the tortoise coordinate $z$, where the conformal infinity is at $z=0$.
We can consider another case of the potential, the right figure of Fig.~\ref{fig:eikonalpotential-gravastar-realspin}.
The modes can be classified into three types.
For $u_c^2<u^2$ and $u_0^2<u^2<u_c^2$, the modes hit the potential one and three times, respectively, as in the previous case.
For $1<u^2<u_0^2$, the mode hits only once near the boundary and cannot be trapped in the potential well around the gravastar surface by tunneling.
We would like to analyze this issue and report on it in near future.

\subsection{Massive Klein-Gordon equation in dS space}
\label{app:dS}

The AdS gravastar analyzed in this note includes the region described by dS spacetime. The wave function of bulk scalar field in the region can be given for dS spacetime, where the exact solution to the wave equation can be obtained.

The metric of four-dimensional dS spacetime can be expressed as
\begin{align}
    ds^2=-f(r)dt^2+f^{-1}(r)dr^2+r^2d\Omega^2,\;\;f(r)=1-r^2,
\end{align}
where we take the dS length as the unit, i.e., $R_\mathrm{dS}=1$.
The massive Klein-Gordon equation $(\Box-m^2)\Phi=0$ reduces to
\begin{align}
\label{eq:dS-waveeq}
    \left[\omega^2-l(l+1)\frac{f}{r^2}-m^2 f+f^2\partial_r^2+\left(\frac{2f}{r}+ff'\right)\partial_r\right]\phi_\ell(r)=0,
\end{align}
by expanding as $\Phi = e^{i\omega t}Y_{\ell m}(\theta ,\phi)\phi_\ell(r)$.
Changing the variable as 
\begin{align}
\phi_\ell=(1-r^2)^{i\omega/2}(r^2)^{\ell/2}\psi_\ell=y^{i\omega/2}(1-y)^{\ell/2}\psi_\ell
\end{align}
with $y=1-r^2$, we find the general solution,
\begin{align}
    &\psi_\ell=Ay^{-i\omega}F(1+a-c,1+b-c;2-c;y)
    +BF(a,b;c;y),\nonumber\\
    &a=\frac{3}{4}+\frac{\ell}{2}-\frac{1}{4}\sqrt{9-4m^2}+\frac{i\omega}{2},\quad
    b=\frac{3}{4}+\frac{\ell}{2}+\frac{1}{4}\sqrt{9-4m^2}+\frac{i\omega}{2},\quad
    c=1+i\omega .
\end{align}
Here $A$ and $B$ are arbitrary constants.
We then have
\begin{align}
\label{eq:dS-gen-sol-horizon}
    \phi_\ell=(1-y)^{\ell/2}\left\{
    Ay^{i\omega/2}F(1+a-c,1+b-c;2-c;y)
    +By^{-i\omega/2}F(a,b;c;y)
    \right\} .
\end{align}
The solution is valid around the cosmological horizon, $y=1-r^2=0$.
Since the hypergeometric function has the value, $F(\cdot,\cdot;\cdot,y)\to 1$ as $y\to0$,
we find
\begin{align}
    \phi_l
    &\sim Ay^{-i\omega/2}+By^{i\omega/2}\nonumber\\
    &\sim A \ 2^{-i\omega}e^{-i\omega z}+B \ 2^{i\omega}e^{i\omega z} \;\; \mathrm{for} \;\;y\to 0.
\end{align}
The tortoise coordinate $z=\int dr/f(r)$ negatively diverges as approaching the cosmological horizon and $(1/2)\ln y\sim z+\ln 2$ around the horizon.
Thus, the terms with coefficients $A$ and $B$ correspond to the ingoing and outgoing modes, respectively.

For the solution around the center, $y=1$, we can use the following property of the hypergeometric function,
\begin{align}
    F(a,b;c;y)
    &=\Gamma_1(a,b,c)F(a,b;a+b+1-c;1-y)\nonumber\\
    &+\Gamma_2(a,b,c)(1-y)^{c-a-b}F(c-a,c-b;1+c-a-b;1-y),\nonumber\\
    \Gamma_1(a,b,c)&:=\frac{\Gamma(c)\Gamma(c-a-b)}{\Gamma(c-a)\Gamma(c-b)},\;\;
    \Gamma_2(a,b,c):=\frac{\Gamma(c)\Gamma(a+b-c)}{\Gamma(a)\Gamma(b)}.
\end{align}
Substituting the equation into the outgoing mode (the mode with the coefficient $B$) in Eq.~\eqref{eq:dS-gen-sol-horizon}, we have
\begin{align}
\label{eq:Bmode-center}
    \phi_\ell|_B
    &=By^{i\omega/2}\bigl\{
    (1-y)^{\ell/2}\Gamma_1(a,b,c)F(a,b;a+b+1-c;1-y)\nonumber\\
    &+(1-y)^{-(\ell+1)/2}\Gamma_2(a,b,c)F(c-a,c-b;1+c-a-b;1-y)
    \bigr\}.
\end{align}
The first and second terms are decaying and growing modes at the center, respectively.
Similarly, for the ingoing mode, we have
\begin{align}
\label{eq:Amode-center}
    &\phi_\ell|_A
    =Ay^{-i\omega/2} \\
    & \times\bigl\{
    (1-y)^{\ell/2}\Gamma_1(1+a-c,1+b-c,2-c)F(1+a-c,1+b-c;1+a+b-c;1-y)\nonumber\\
    & \quad \quad +(1-y)^{-(\ell+1)/2}\Gamma_2(1+a-c,1+b-c,2-c)F(1-a,1-b;1+c-a-b;1-y)
    \bigr\} . \nonumber 
\end{align}
The first and second terms are decaying and growing modes at the center, respectively.
We would like to impose the regularity as the boundary condition at the center.
It requires vanishing of the growing part of $\phi_\ell|_A+\phi_\ell|_B$ at $y=1$.
The condition for the coefficients reduces to
\begin{align}
\label{eq:regularity-BA}
    \frac{B}{A}
    &=-\frac{\Gamma_2(1+a-c,1+b-c,2-c)}{\Gamma_2(a,b,c)}\nonumber\\
    &=-\frac{\Gamma(a)\Gamma(b)\Gamma(2-c)}{\Gamma(1+a-c)\Gamma(1+b-c)\Gamma(c)}.
\end{align}
The condition is satisfied for the coefficients set in \eqref{eq:psicoeff} for \eqref{eq:psi}.
In our gravastar case, if the junction radius is nearly equal to the cosmological horizon radius, $y=0$, this ratio of the coefficients would coincide with that of the ingoing and outgoing modes just outside the gravastar.

\bibliographystyle{JHEP}
\bibliography{PhotonSphere}

\end{document}